\documentclass[aps,prd,twocolumn,nofootinbib,superscriptaddress]{revtex4-1}
\usepackage{amstext}
\usepackage{amssymb}
\usepackage{graphicx}
\usepackage{subfigure}
\usepackage{color}
\usepackage{natbib}
\usepackage{url}
\usepackage[colorlinks=true]{hyperref}
\usepackage{acronym}

\def\msun{M_{\odot}}
\def\mbh{m_{\rm BH}}
\def\mns{m_{\rm NS}}
\def\chibh{\chi_{\rm BH}}
\def\chins{\chi_{\rm NS}}

\def\chiperp{\chi^{\perp}_{\rm BH}}
\def\chipar{\chi^{\parallel}_{\rm BH}}

\begin{document}

\title{Impact of precession on aligned-spin searches for neutron-star--black-hole binaries}

\author{Tito \surname{Dal Canton}}
\email{tito.dalcanton@aei.mpg.de}
\affiliation{Max-Planck-Institut f\"ur Gravitationsphysik, Callinstrasse 38,
D-30167 Hannover, Germany}
\author{Andrew P. Lundgren}
\email{andrew.lundgren@ligo.org}
\affiliation{Max-Planck-Institut f\"ur Gravitationsphysik, Callinstrasse 38,
D-30167 Hannover, Germany}
\author{Alex B. Nielsen}
\email{alex.nielsen@ligo.org}
\affiliation{Max-Planck-Institut f\"ur Gravitationsphysik, Callinstrasse 38,
D-30167 Hannover, Germany}

\date{\today}

\begin{abstract}
The inclusion of aligned-spin effects in gravitational-wave search pipelines for
neutron-star--black-hole binary coalescence has been shown to increase the
astrophysical reach with respect to search methods where spins are neglected
completely, under astrophysically reasonable assumptions about black-hole
spins. However, theoretical considerations and population synthesis models
suggest that many of these binaries may have a significant misalignment between
the black-hole spin and the orbital angular momentum, which could lead to
precession of the orbital plane during the inspiral and a consequent loss in
detection efficiency if precession is ignored. This work explores the effect of
spin misalignment on a search pipeline that completely neglects spin effects and on a
recently-developed pipeline that only includes aligned-spin effects. Using
synthetic but realistic data, which could reasonably represent the first
scientific runs of advanced-LIGO detectors, the relative sensitivities of both
pipelines are shown for different assumptions about black-hole spin magnitude
and alignment with the orbital angular momentum.  Despite the inclusion of
aligned-spin effects, the loss in signal-to-noise ratio due to precession can
be as large as $40\%$, but this has a limited impact on the overall detection
rate: even if precession is a predominant feature of neutron-star--black-hole
binaries, an aligned-spin search pipeline can still detect at least half of the signals
compared to an idealized generic precessing search pipeline.
\end{abstract}

\maketitle

\acrodef{GW}{gravitational wave}
\acrodef{BH}{black hole}
\acrodef{NS}{neutron star}
\acrodef{NSBH}{neutron-star--black-hole}
\acrodef{PSD}{power spectral density}
\acrodef{MECO}{minimum-energy circular orbit}
\acrodef{ROC}{receiver operating characteristic}
\acrodef{SNR}{signal-to-noise ratio}
\acrodef{NSB}{nonspinning bank}
\acrodef{ASB}{aligned-spin bank}
\acrodef{ISCO}{innermost stable circular orbit}
\acrodef{pN}{post-Newtonian}

\section{Introduction}
\label{sec:intro}
Astrophysical \acp{BH} can be simply described by their mass and spin angular
momentum. A large number of \ac{BH} properties can be derived from knowing just
these two values. Searching for faint \ac{GW} signals from \acp{BH} in
coalescing binary systems using ground-based interferometers such as advanced
LIGO \cite{Harry:2010zz}, advanced Virgo \cite{TheVirgo:2014hva} and KAGRA
\cite{Somiya:2011np} requires a bank of potential inspiral signals (templates)
over a range of possible parameter values. Such template banks have
traditionally been built ranging over possible \ac{BH} masses, but in most
cases assuming the absence of spin (see e.g.~\cite{Abbott:2009tt,
Abadie:2010yb, Abadie:2011kd, Colaboration:2011np, oai:arXiv.org:1209.6533}).
This was recently extended to enable searches for gravitational radiation from
coalescing \ac{NSBH} binaries which include aligned-spin effects
\cite{Brown:2012qf, PhysRevD.89.084041, pycbcpaper}. Under reasonable
distributions of binary configurations and the assumption that the \ac{BH} spin
and orbital angular momentum are aligned, including spin effects has been shown
to improve the sensitivity of the search method with respect to methods
neglecting the effect of spin altogether \cite{pycbcpaper}.

However, \ac{NSBH} binaries are generally expected to have a \ac{BH} spin
misaligned with respect to the orbital angular momentum. The formation of
\ac{NSBH} binaries coalescing in a time useful for detection seems to require a
dissipative common-envelope phase \cite{O'Shaughnessy:2005qc, Dominik:2012kk},
which tends to align the angular momenta and spin up the \ac{BH}, followed by
the supernova explosion of the smaller object, which imparts a significant kick
to the resulting \ac{NS}. Binaries surviving the second supernova explosion
turn out to have a \ac{NS} kick that effectively tilts the orbital plane with
respect to the \ac{BH} spin \cite{Kalogera:1999tq}.

Nevertheless, tilting the orbit by large angles turns out to be very hard;
simulations suggest that most \ac{NSBH} binaries will generally have a
misalignment angle smaller than about 60 degrees, with a large fraction having
an angle smaller than 45 degrees \cite{PhysRevD.69.102002, Belczynski:2007xg,
Fragos:2010tm}. Unfortunately, X-ray studies of \ac{BH} spin values are not
able to measure directly the alignment of the spin; instead, they rely on the
assumption that the accretion disk should align with the spin, and
observational constraints on spin magnitudes can thus be influenced by
misalignment \cite{Fragile:2009vu}.

A misalignment between the \ac{BH} spin and the orbital angular momentum breaks
a symmetry of the system and leads to a precessing orbital plane. The
time-varying orientation of the orbital angular momentum then causes a
characteristic phase and amplitude modulation of the chirping \ac{GW} signal
observed on Earth \cite{Apostolatos:1994mx, BLO} which is absent in a
nonprecessing waveform.

Including the effect of precession in a \ac{GW} search pipeline has been
attempted before, but it did not yield a significant gain in sensitivity
\cite{VanDenBroeck:2009gd, Abbott:2007ai}. The best strategy for including such
effects is still unknown, as are the resources required by a fully-precessing
search pipeline. Thus, even if the spin-orbit misalignment may not be very
large, it is important to assess its impact on a search pipeline that
completely neglects precession.  In light of the results of
\cite{Abbott:2007ai} it is possible that an aligned-spin search pipeline
performs no better than a nonspinning one. It is therefore important to
determine how precession affects the sensitivity of including aligned-spin
effects over a cheaper method that completely neglects spin. This problem has
been recently investigated using bank simulations and assuming the final design
sensitivity of advanced LIGO interferometers \cite{PhysRevD.89.084041,
Harry:2013tca}.  Here the investigation is extended, in particular it is
applied to realistic data rather than idealized noise and assuming an
``early-advanced-LIGO'' sensitivity curve, which more realistically represents
the first (2015) scientific runs of advanced LIGO. The curve is available at
\cite{LIGO-T1200307-v4} and plotted in Figure 1 of \cite{pycbcpaper} (black
solid line).

Using bank simulations, we first study the loss in \ac{SNR} imparted by
precession when using template banks which (i) neglect spin altogether and (ii)
only include aligned-spin effects. For both cases we explore the dependency of
the loss on the parameters which mainly affect precession, i.e.~the magnitude
and tilt of the \ac{BH} spin. We also use the bank simulations to estimate the
loss of detections produced by considering aligned-spin effects but neglecting
precession. Then, using the approach described in \cite{pycbcpaper}, we search
for a population of simulated precessing \ac{NSBH} binaries in synthetic data
and we present the sensitivity of the search pipeline for different possible
distributions of \ac{BH} spins. The synthetic data we analyze consist of 60
days of real data (of which about 25 are useful for the analysis) from the
sixth scientific run of the initial LIGO Hanford and Livingston
interferometers. However, the \ac{PSD} of the data is modified to resemble the
aforementioned early-advanced-LIGO sensitivity curve, in order to reasonably
represent data expected from the first scientific runs of advanced LIGO.

Although the merger and ringdown parts of \ac{NSBH} signals happen within the
bandwidth of the detectors for some regions of the \ac{NSBH} parameter space,
particularly anti-aligned systems, we neglect them in this work and focus
instead on purely precessional effects, which we expect to be somewhat
complementary. We reserve a comprehensive study of merger and ringdown effects
to a future article.

The paper is organized as follows. In section \ref{sec:effectualness} we
present bank simulations showing the effectualness of a nonspinning template
bank for precessing \ac{NSBH} signals and of a spinning but nonprecessing
bank. In section \ref{sec:search} we present the results of running our search
pipeline, using the above banks, on a stretch of realistic data containing
simulated precessing \ac{NSBH} binaries. Section \ref{sec:conclusion}
summarizes conclusions and future work.

\section{Effect of precession on \ac{SNR} loss}
\label{sec:effectualness}

The first step in studying the effect of precession is calculating the fitting
factor between the population of precessing signals we want to observe and the
template banks we intend to use, which is equivalent to asking how good, on
average, our bank is at recovering the signals we target. This can be done via
bank simulations. For a full description of the basic ideas behind template
banks and bank simulations, as well as more details of the banks we use in this
section, see \cite{pycbcpaper}.

The source population we are interested in contains \ac{NSBH} binaries with
\ac{BH} mass $\mbh$ between $3\,\msun$ and $15\,\msun$ and \ac{NS} mass $\mns$
between $1\,\msun$ and $3\,\msun$. The dimensionless \ac{BH} spin has magnitude
uniformly distributed within the bounds imposed by the Kerr solution, $0 \leq
\chibh \leq 1$. The tilt angle $\vartheta$ of the \ac{BH} spin with respect to
the orbital angular momentum is defined by a uniform distribution of $\kappa
\equiv \cos \vartheta$ in the range $\pm 1$, such that $\hat\chibh$ is
uniformly distributed on the sphere.  Although the resulting spin distribution
is a reasonable choice if no information is available about \ac{BH} spins, a
number of studies and observations exist suggesting that $\chibh$ may be large
\cite{O'Shaughnessy:2005qc, McClintock:2011zq}. At the same time, models based
on population synthesis suggest that $\vartheta$ is likely peaked at 0, with
only a small fraction of \ac{NSBH} binaries having $\vartheta > 60$ degrees
\cite{PhysRevD.69.102002, Belczynski:2007xg}. In other words, there is a small
region of the ($\chibh,\vartheta$) plane which may be much more common
astrophysically than the rest. The \ac{NS} spin is known to have a small effect
on a search for \ac{NSBH} coalescence \cite{pycbcpaper}. Nevertheless we take a
uniform distribution of its magnitude in the range $0 \leq \chins \leq 0.05$,
which should include the fastest spinning \acp{NS} observed in compact binaries
(see e.g.~\cite{Damour:2012yf}). The distribution of $\hat\chins$ is also
uniform on the unit sphere.

The \ac{NSB} we use here covers \ac{BH} masses between $3\,\msun$ and
$15\,\msun$ and \ac{NS} masses between $1\,\msun$ and the smallest of either
the equal-mass boundary or the line corresponding to a total mass of
$18\,\msun$. The \ac{ASB} covers the same \ac{BH} masses, \ac{NS} masses
between $1\,\msun$ and $3\,\msun$ only, dimensionless \ac{BH} spin projection
along the orbital angular momentum $\vec\chi_{\rm BH} \cdot \hat L \in [-1,1]$
and \ac{NS} spin projection $\vec\chi_{\rm NS} \cdot \hat L \in [-0.4,0.4]$.
The \ac{NSB} and \ac{ASB} contain $\sim 2.8 \times 10^4$ and $\sim 1.5 \times
10^5$ templates respectively; including aligned spin leads to a $\sim 5\times$
larger computing cost.  Template waveforms are computed in the frequency domain
using the TaylorF2 approximant.  Their inspiral phasing contains orbital terms
up to 3.5 \ac{pN} order and spin-orbit terms up to 2.5 \ac{pN}. Waveforms start
at 30 Hz and terminate at the frequency of the \ac{ISCO}.

\begin{figure*}
	\centering
    \includegraphics[width=2\columnwidth]{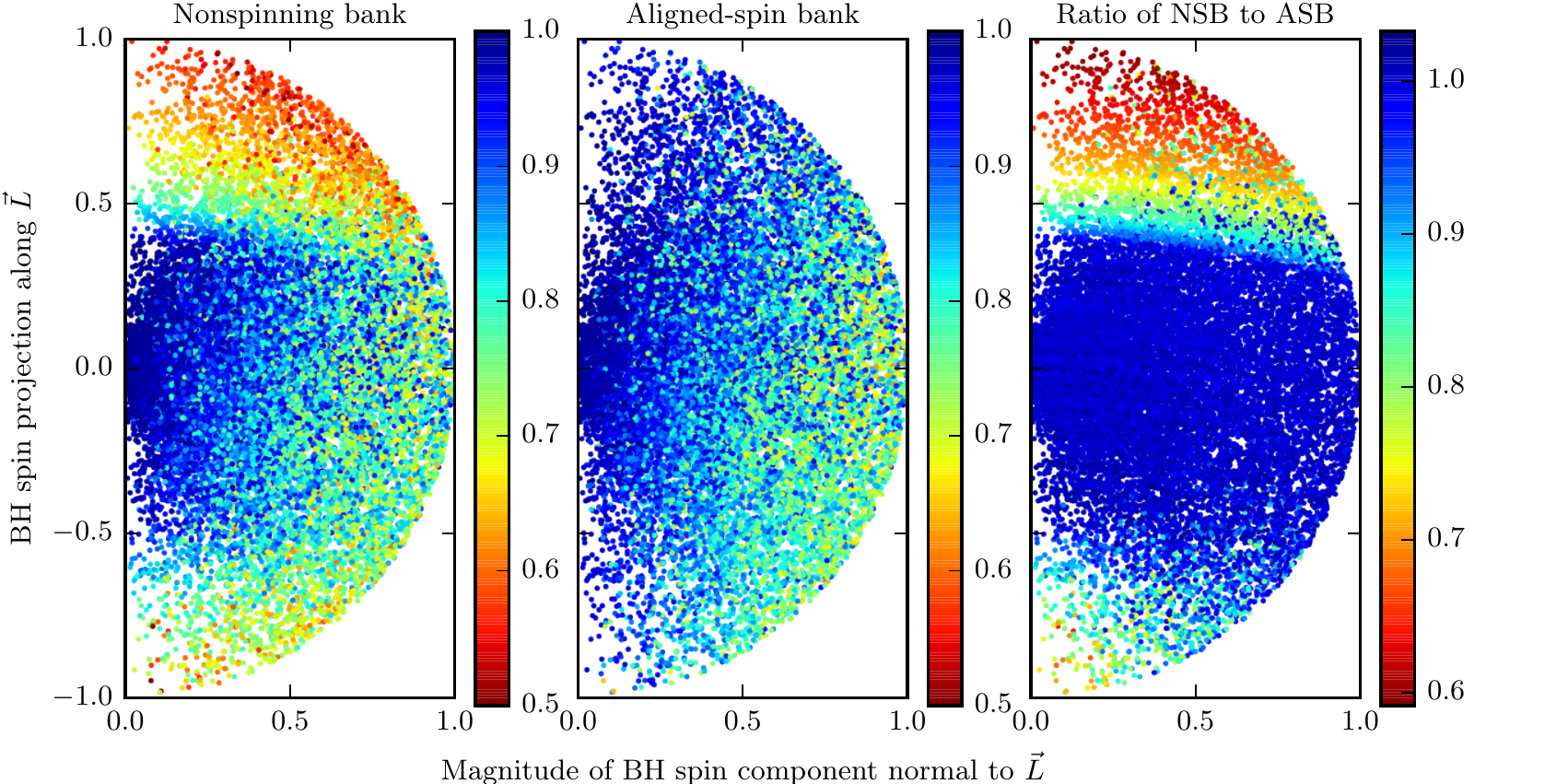}
    \caption{Fitting factor of our nonspinning bank (left), aligned-spin
    bank (middle) and their ratio (right) for a population of 20000 precessing
    \ac{NSBH} binaries with fixed masses $\mbh = 7.8 \msun$, $\mns = 1.35 \msun$.
    The horizontal axis is the amount of \ac{BH} spin orthogonal to the orbital
    angular momentum, while the vertical axis is its projection along the
    orbital angular momentum; thus, the origin corresponds to nonspinning
    signals and the vertical axis corresponds to the aligned-spin case. The
    spin magnitude and orientation on the sphere are both uniformly
    distributed.}
	\label{fig:banksim_fix_mass}
\end{figure*}

In order to separate the effect of spin from mass-related issues of the banks,
we first restrict our attention to binaries with fixed masses $\mbh = 7.8
\msun$ and $\mns = 1.35 \msun$, which represent typical mass values for
\acp{BH} and \acp{NS} in binaries \cite{Ozel:2010su, Ozel:2012ax}. This allows
us to study the fitting factor of the \ac{NSB} and \ac{ASB} as a function of
the \ac{BH} spin parameters only and with high statistics. The result of this
first set of simulations is shown in Figure \ref{fig:banksim_fix_mass}, which
displays the fitting factor as a function of the amount of \ac{BH} spin
orthogonal to the orbital angular momentum and parallel to it,
\begin{eqnarray}
    \chiperp &\equiv& ||\vec\chibh - (\vec\chibh \cdot \hat L) \hat L|| \\
    \chipar &\equiv& \vec\chibh \cdot \hat L.
\end{eqnarray}
The behavior of the two banks allows us to roughly divide this space in three
approximately defined regions.

The first region is defined by low values of the \ac{BH} spin magnitude,
approximately $\chibh \lesssim 0.4$.  Not surprisingly, both banks perform well
here. The spin-orbit terms in the signal waveforms are small enough that the
consequent dephasing can be accommodated by a small bias in the symmetric mass
ratio $\eta$ \cite{PhysRevD.87.024035} allowing the \ac{NSB} to recover signals
well. Moreover, even if the spin is tilted, its magnitude is too small for
precession to induce significant modulation.

The second region is defined by a small perpendicular component of the \ac{BH}
spin but a large parallel component (positive or negative), i.e.~$\chiperp
\lesssim 0.5$ and either $\chipar \gtrsim 0.4$ or $\chipar \lesssim -0.5$.
Here the \ac{NSB} has a sudden and severe loss of effectualness, while the
\ac{ASB} performs significantly better.  This loss happens because the
spin-orbit terms in the waveform phase acquire their most extreme values, so
neglecting them causes the largest possible dephasing.  The resulting bias in
$\eta$ when ignoring spin is too large to fit into the \ac{NSB}: unphysical
templates with $\eta > 1/4$ would be required to recover positively-aligned
signals and templates with $\mns < 1$ would be needed to match
negatively-aligned ones.  In other words, signals in this region ``fall off''
the \ac{NSB} \cite{pycbcpaper}.  However, the signal modulation due to
precession is still small in this region, so the \ac{ASB} is able to recover
almost all the signal power.

The third region is the highly-precessing case, roughly identified by $\chiperp
\gtrsim 0.5$. The main features here are the very similar performance of both
banks and the significant spread of the fitting factor with respect to the
other two regions, with some sources being recovered well and others poorly.
The former effect happens because spin-orbit terms are small, so the \ac{NSB}
is still able to compensate them by using templates with a biased symmetric
mass ratio.  At the same time, the modulation induced by precession is strong
and both banks are equally unable to recover the power going into the
modulation sidebands. This suggests that the effects of the modulation and of
$\chipar$ effectively decouple, as predicted for example in
\cite{Lundgren:2013jla, Schmidt:2014iyl}.

\begin{figure}
    \centering
    \includegraphics[width=\columnwidth]{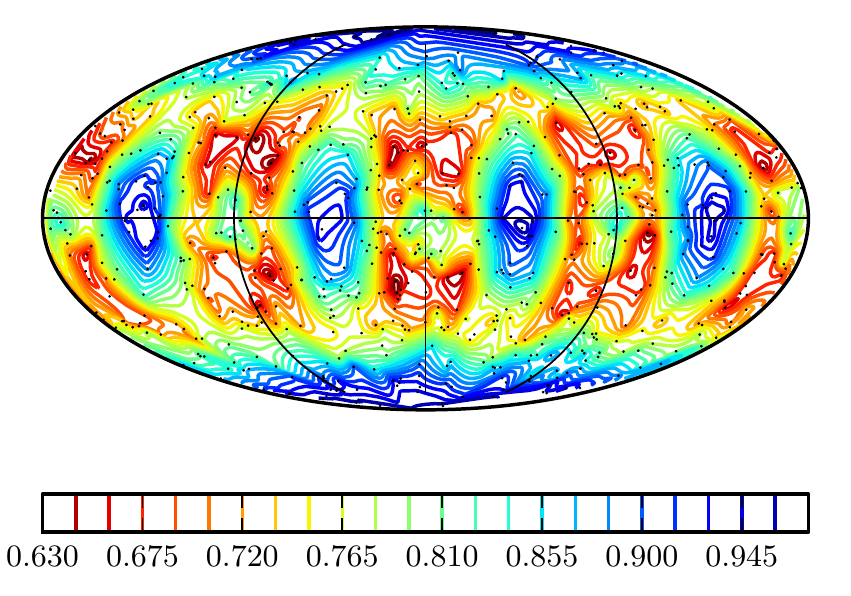}
    \caption{Fitting factor of the aligned-spin bank for
    strongly-precessing \ac{NSBH} binaries as a function of the orientation of the
    total angular momentum with respect to the detector. The south and north
    poles of the projection correspond to face-on and face-off orientations,
    while the equatorial line is the edge-on case. Sources have fixed masses
    $\mbh = 7.8 \msun$ and $\mns = 1.35 \msun$ and spin parameters $\chibh >
    0.8$ and $|\kappa| < 0.2$. Each source is shown as a black dot.}
    \label{fig:orientation}
\end{figure}

The second effect visible in this region, i.e.~the scatter of the fitting
factor, is due to the different possible orientations of the total angular
momentum $\vec J$ with respect to the detector, as demonstrated in Figure
\ref{fig:orientation}.
In fact, the orientation changes the fraction of signal power in the modulation
sidebands, which nonprecessing templates cannot recover.  In particular,
face-on and face-off binaries almost look like nonprecessing systems when
viewed from the detector, so their waveforms have a smaller modulation, and
thus larger fitting factor, than edge-on systems.  Note that edge-on systems
can also have good fitting factors at four particular orientations of $\vec J$;
however, those orientations produce quiet signals at the detector.  Because of
the almost-linear polarization of their radiation, edge-on systems are in fact
generally quieter than face-on or face-off ones at the same distance. Thus,
although recovering them is challenging even with the \ac{ASB}, they are also
the least detectable even in the ideal case.

One can therefore ask whether the poor performance of the \ac{ASB} in the
high-precession case is really affecting the overall sensitive volume to a
population of binaries with uniform spin and angle distribution as we assume
here. The bank simulation enables a rough estimate of the sensitive volume of
an aligned-spin search pipeline relative to a hypothetical ideal generic
precessing pipeline:
\begin{equation}
    \mathcal{V} \equiv \frac{V_{\rm ASB}}{V_{\rm prec}} \simeq \frac{\sum_i (m_i \rho_i)^3}{m^3_{\rm prec} \sum_i \rho^3_i}.
    \label{eq:vestim}
\end{equation}
Here the sums are over the simulated signals, $m_i$ is the fitting factor of
signal $i$ obtained from the bank simulation, $\rho_i$ is the optimal \ac{SNR}
of signal $i$ at a fixed reference distance and $m_{\rm prec} = 98.5\%$ is a
guess for the average fitting factor of the hypothetical precessing bank.  This
estimate does not include the likely increased false-alarm background of the
precessing pipeline, which would increase $\mathcal{V}$. Evaluating this increase
requires constructing a generic precessing template bank, which is an open
problem and is outside the scope of this paper.  The estimation also neglects
the effect of signal-based vetoes, which would instead reduce $\mathcal{V}$.
The effect of vetoes, coincidence and realistic data can be included by
running a full aligned-spin search pipeline, which will be described in Section
\ref{sec:search}. With this caveat in mind, the fixed-mass bank simulation
gives, for no restriction on spin,
\begin{equation}
    \mathcal{V}^{\rm all}_{\rm FM} \simeq 81\%.
\end{equation}
If we restrict our attention to highly-precessing binaries, say with $\chibh >
0.7$ and $\frac{1}{4}\pi < \vartheta < \frac{3}{4}\pi$, we obtain instead
\begin{equation}
    \mathcal{V}^{\rm HP}_{\rm FM} \simeq 61\%.
    \label{eq:vhpfm}
\end{equation}

\begin{figure*}
    \centering
    \includegraphics[width=2\columnwidth]{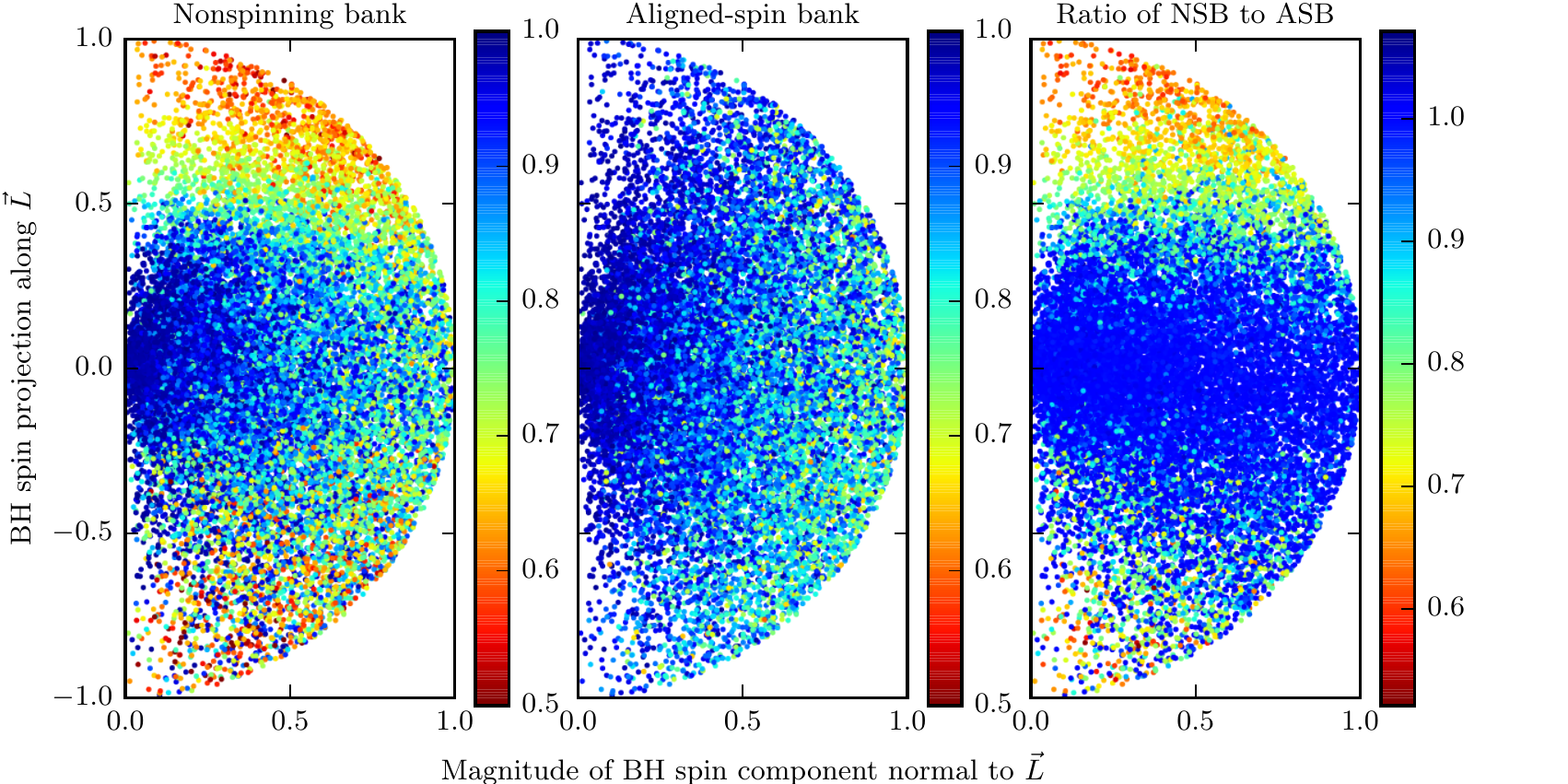}
    \caption{Fitting factor of our nonspinning bank (left), aligned-spin
    bank (middle) and their ratio (right) for a population of 20000 precessing
    \ac{NSBH} binaries with the mass distribution described in the text.
    Compare with Figure \ref{fig:banksim_fix_mass}.}
    \label{fig:banksim_var_mass}
\end{figure*}

We now repeat the bank simulation for the full distribution of masses described
earlier in this section. The spin parameters are distributed as before. Figure
\ref{fig:banksim_var_mass} shows the results and it can be seen that the match
variation is qualitatively consistent with the fixed-mass case.
A noticeable difference, however, is the much worse mismatch of the \ac{NSB}
for some of the strongly-spinning and anti-aligned systems (lower region of
first and last plots). These happen to be heavy systems, with \ac{BH} mass
larger than about $11\,\msun$. Their poor match is not due to precession but
again to the bias in symmetric mass ratio produced when attempting to recover
spinning signals with zero-spin templates. In this case the spin is
anti-aligned, so the bias is negative: signals are well matched by templates
associated with a lighter \ac{NS} and a heavier \ac{BH}. For signals with small
enough mass, the bias is within the mass space covered by the \ac{NSB} and a
good fitting factor is obtained. Signals with negatively-spinning \acp{BH}
heavier than $\sim 11\,\msun$, however, ``fall off'' the high-\ac{BH}-mass edge
of the \ac{NSB}; templates with \ac{BH} masses higher than $15\,\msun$ would be
needed to recover these signals. This effect is less dramatic in the fixed-mass
bank simulation because the masses are far from the boundary of the bank.

As done for the fixed-mass bank simulation, we can again estimate the loss of
detections of the aligned-spin pipeline relative to a hypothetical generic
precessing pipeline. Using Eq.~\ref{eq:vestim} with the result of the
varying-mass bank simulation yields
\begin{equation}
    \mathcal{V}^{\rm all}_{\rm VM} \simeq 86\%.
\end{equation}
High-mass systems have a larger weight in this estimate due to the higher
intrinsic loudness of their signals and they also exhibit the largest
precession effects due to the high mass ratio. $\mathcal{V}_{\rm VM}$ may
therefore overestimate precession effects, but nevertheless the resulting loss
is still quite small. The relative volume for varying-mass, highly-precessing
systems ($\chibh > 0.7$ and $\frac{1}{4}\pi < \vartheta < \frac{3}{4}\pi$) is
\begin{equation}
    \mathcal{V}^{\rm HP}_{\rm VM} \simeq 71\%.
    \label{eq:vhpvm}
\end{equation}

These two first exercises show that: (i) the \ac{ASB} is much better exactly
where population synthesis models suggest the majority of signals will be
($\chibh \to 1$ and $\vartheta \to 0$); (ii) anti-aligned systems ($\chibh \to
1$ and $\vartheta \to \pi$) give rise to the worst matches in the \ac{NSB}, but
this problem could be alleviated by adding heavier-\ac{BH} templates to the
\ac{NSB}; (iii) both banks show a comparable inefficiency with strong
precession, i.e.~near maximal spin misalignments ($\vartheta \to \pi/2$), but
this effect is smaller than the loss imparted by neglecting spin-orbit terms
and only happens for some orientations of the total angular momentum; (iv) even
if a generic precessing template bank is not yet available, when taking into
account the orientation-dependent intrinsic loudness of signals, the poor
performance of the \ac{ASB} with strong precession seems to reduce the
detection rate by a few tens of percent at most. This result is compatible with
similar existing studies \cite{PhysRevD.89.084041, Harry:2013tca}. However, the
estimate only considers the loss of \ac{SNR} and obviously needs to be
evaluated more precisely by running a full, realistic search pipeline.

\section{Effect of precession on a realistic search pipeline}
\label{sec:search}
The overall efficiency of a search is determined primarily by two features.
The first is the background of false alarms generated by the template bank due
to the detector noise. It has been shown in our previous study that the
increase in background due to using the larger \ac{ASB} is not considerable
relative to the \ac{NSB} when the reweighted \ac{SNR} is used as a ranking
statistic \cite{pycbcpaper}. The second feature is the ability of the waveforms
in the chosen template bank to match the targeted signals; we studied that in
the previous section. Based on these results, we expect a search using the
\ac{ASB} to perform at least as well as the \ac{NSB}, and to significantly
outperform it for a population of almost-aligned systems. However, it remains
to be checked whether coincidence between the detectors and the inclusion of
the $\chi^2$ veto alter the result significantly in the presence of precession.
As a final result, we also want to compare the search sensitivity at fixed
false-alarm rate.

In this section, we thus apply the search method described in
\cite{pycbcpaper}, with identical parameters and data, to the fixed-mass and
variable-mass precessing signal populations described in the previous section.
We recall that this is a matched-filter, exact-match-coincidence pipeline based
on the PyCBC toolkit, with settings similar to what used in the last
initial-LIGO searches. The data we analyze are constructed using two months of
data from the sixth science run of the Hanford and Livingston LIGO detectors
and modifying their \ac{PSD} to resemble the early-advanced-LIGO sensitivity
curve we assume in this work.

\begin{figure}
    \centering
    \includegraphics[width=\columnwidth]{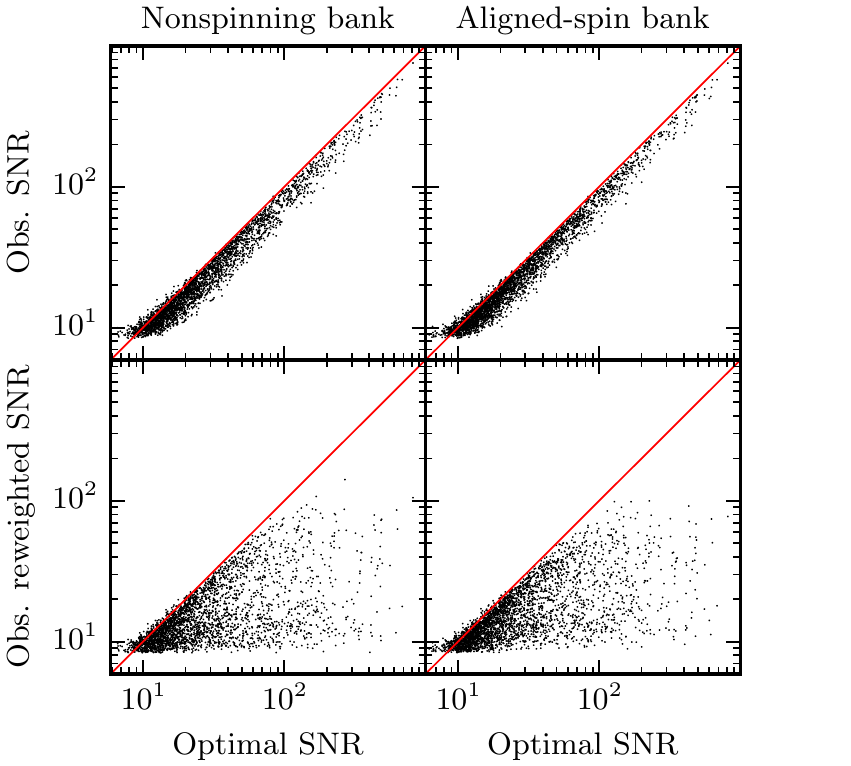}
    \caption{Combined \ac{SNR} and reweighted \ac{SNR} observed by our search
    pipeline versus the optimal \ac{SNR} one would achieve by using templates
    perfectly matched to each simulated system. The source population has fixed
    masses $\mbh = 7.8 \msun$, $\mns = 1.35 \msun$. There is no clear
    difference between the left and right plots because relatively few points
    have large and almost aligned spins.}
    \label{fig:prediction1}
\end{figure}

\begin{figure*}
    \centering
    \includegraphics[width=2\columnwidth]{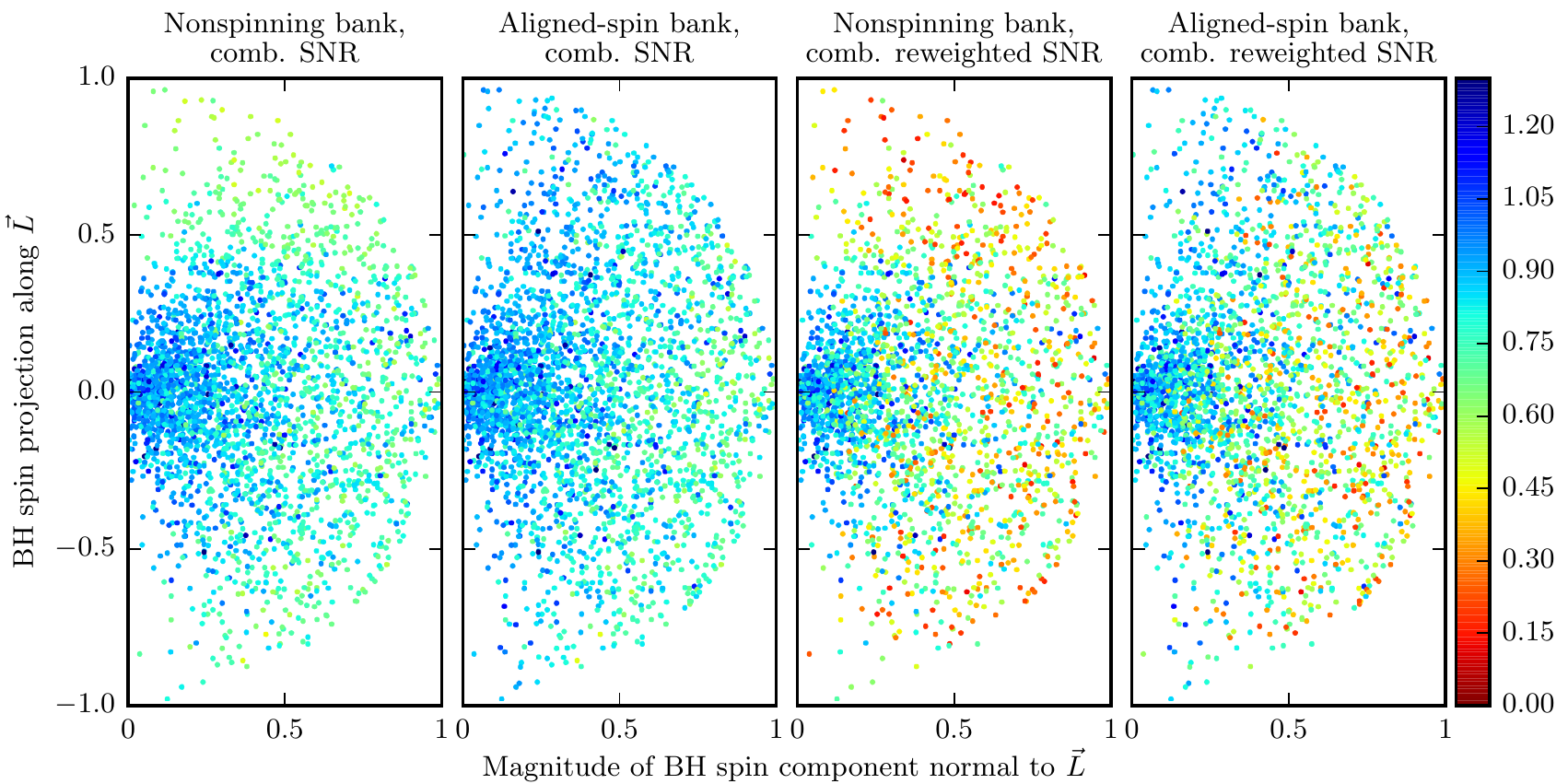}
    \caption{Ratio between combined \ac{SNR} or combined reweighted \ac{SNR}
    observed by our search pipeline and the optimal \ac{SNR} one would achieve
    by using templates perfectly matched to each simulated system. The source
    population has fixed masses $\mbh = 7.8 \msun$, $\mns = 1.35 \msun$.  In
    the reweighted \ac{SNR} plots, only systems with an observed combined
    \ac{SNR} smaller than 50 are shown, because for louder systems the
    observed reweighted \ac{SNR} is much smaller than the optimal \ac{SNR}.}
    \label{fig:prediction2}
\end{figure*}

Similarly to what was done in \cite{pycbcpaper}, we first compare, for each
simulated system, the combined \ac{SNR} and combined reweighted \ac{SNR}
observed by the pipeline to the optimal combined \ac{SNR}. The latter is
calculated by simulating the system's waveform and using it as its own optimal
template\footnote{Note that, with precessing signals, waveforms observed by
different detectors are not simply related by phase rotations and amplitude
scalings, but can be qualitatively different due to the variable relative
orientations between the system and the detectors. Therefore, the optimal
template in one detector is, in general, not optimal for another detector.}.
The comparison for the fixed-mass population is shown in Figures
\ref{fig:prediction1} and \ref{fig:prediction2}; results for the varying-mass
systems are consistent with the fixed-mass case and are not shown.
In Figure \ref{fig:prediction1} it can be seen that many systems have a large
loss of both ranking statistics. The reweighted \ac{SNR} is particularly affected,
which can be explained by the $\chi^2$ veto penalizing strongly precessing
signals in both banks. As the signals shown in Figure \ref{fig:prediction1}
have no constraints on spin parameters, and relatively few points have high and
quasi-aligned systems, the advantage of the \ac{ASB} is not clearly visible in
these plots. Figure \ref{fig:prediction2} shows instead the observed fraction
of optimal \ac{SNR} across the spin parameter space and can be compared with
Figure \ref{fig:banksim_fix_mass}. The optimal \ac{SNR} is not a good predictor
of the reweighted \ac{SNR} when signals become too loud, because in that
regime even a small residual mismatch causes the $\chi^2$ statistic to become
very large. In order to make Figure \ref{fig:prediction2} more clear, thus, the
reweighted \ac{SNR} plots exclude signals with observed combined \ac{SNR}
larger than 50.  For both statistics the results are consistent with
Sec.~\ref{sec:effectualness}: the \ac{ASB} performance is superior to the \ac{NSB}
for large spin and tilt angle close to 0 or 180 degrees, while both banks
perform similarly for small spin or tilt angle around 90 degrees. In
particular, looking at the top of the plots, it can be seen that many high-spin
and small-tilt signals are not detected at all by the \ac{NSB}.  Thus, the
$\chi^2$ veto and exact-match coincidence preserve the features of the banks
described in Sec.~\ref{sec:effectualness}.

\begin{figure*}
    \centering
    \includegraphics[width=2\columnwidth]{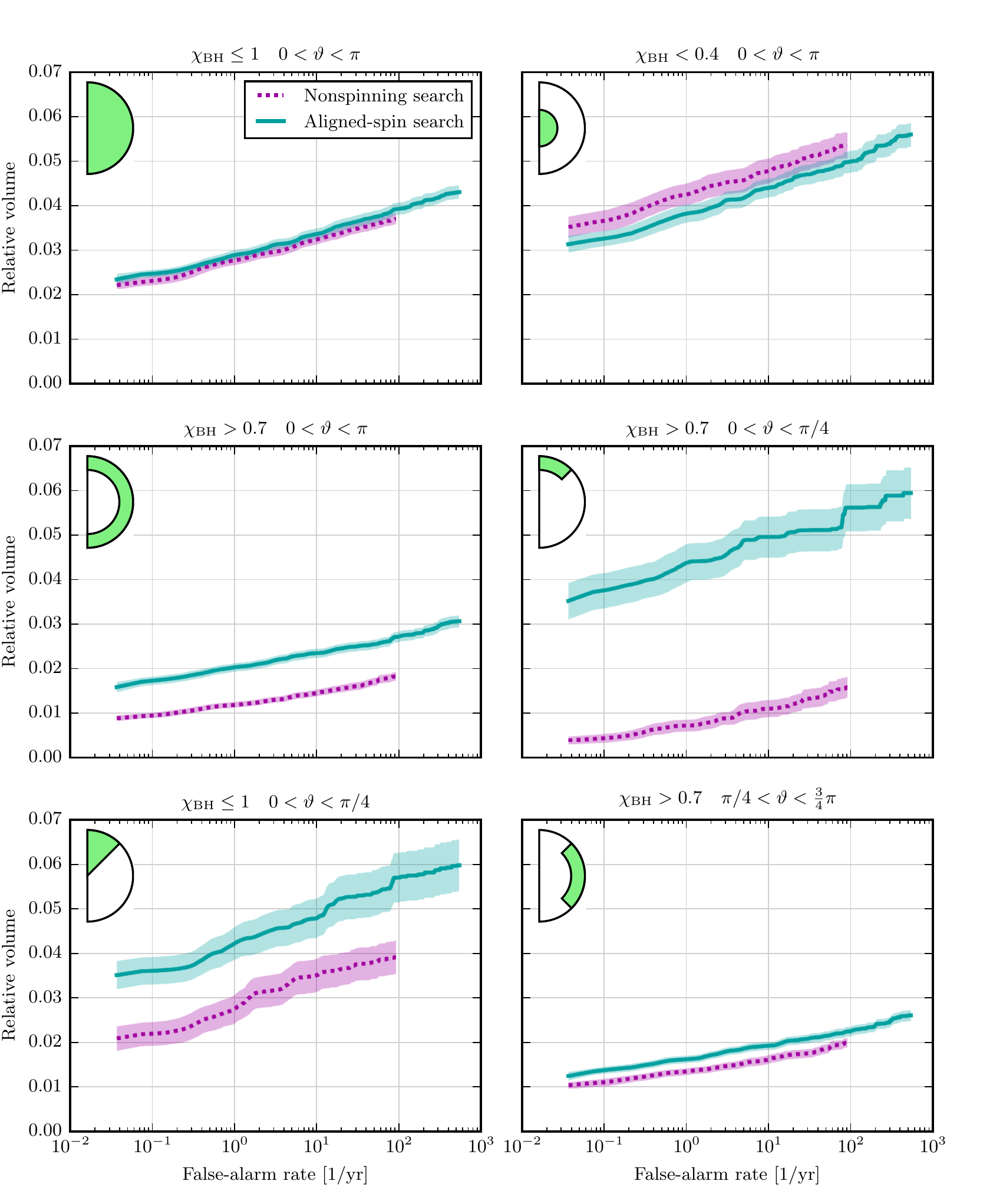}
    \caption{\acp{ROC} associated with nonspinning and spinning search pipelines
    (dashed and solid lines respectively) observing populations of \ac{NSBH}
    binaries with fixed masses $\mbh = 7.8 \msun$, $\mns = 1.35 \msun$ and
    different constraints on the \ac{BH} spin parameters $(\chiperp,\chipar)$
    (visualized in the insets). The lighter bands show the 68\% error intervals
    estimated by constructing each curve 100 times from different combinations
    of the available data.}
    \label{fig:rocfixmass}
\end{figure*}

As a final step, we turn our attention to \ac{ROC} curves, showing the
sensitive volume of the pipeline (proportional to the detection rate) as a
function of the false-alarm rate\footnote{The exact definition of \acp{ROC}
used here is equation 15 of \cite{pycbcpaper}.}. \acp{ROC} for fixed masses are
shown in Figure \ref{fig:rocfixmass}, where different plots compare the
pipelines using the \ac{NSB} (dashed lines) and \ac{ASB} (solid lines) for
different cuts on the \ac{BH} spin magnitude $\chibh$ and tilt angle
$\vartheta$. Error intervals represent the standard deviation of 100
realizations of each curve, each constructed from a random selection of half of
the background and half of the simulated signals. The first plot contains
systems with no restrictions on spin parameters and in this case the \ac{ASB}
gives a slightly larger sensitivity, although the difference is within the
error intervals. The second plot corresponds to weakly-spinning \acp{BH} with
no restriction on tilt angles.  Such a population would be almost ``tuned'' to
the \ac{NSB} and in fact the \ac{ASB} has a slightly lower sensitivity due to
the larger background.  The difference is nevertheless still comparable to the
error intervals. The next three plots assume respectively large spins, large
spins and small tilt angles, and small tilt angles only.  In this case the
\ac{ASB} has a clear advantage, with a sensitivity between 50\% and one order
of magnitude larger.  If the existing population synthesis models and spin
measurements are assumed, the most realistic scenario would lie somewhere
between plots 4 and 5. We can also see from Figure \ref{fig:rocfixmass} that
the sensitivity of the \ac{ASB} remains good for most choices of cuts on spin
parameters, while the sensitivity of the nonspinning pipeline drops
significantly when $\chibh > 0.7$.  Nevertheless, if a large fraction of
sources has strong and misaligned spins (plots 3 and 6) the detection rate
drops noticeably also with the aligned-spin pipeline.

\begin{figure*}
    \centering
    \includegraphics[width=2\columnwidth]{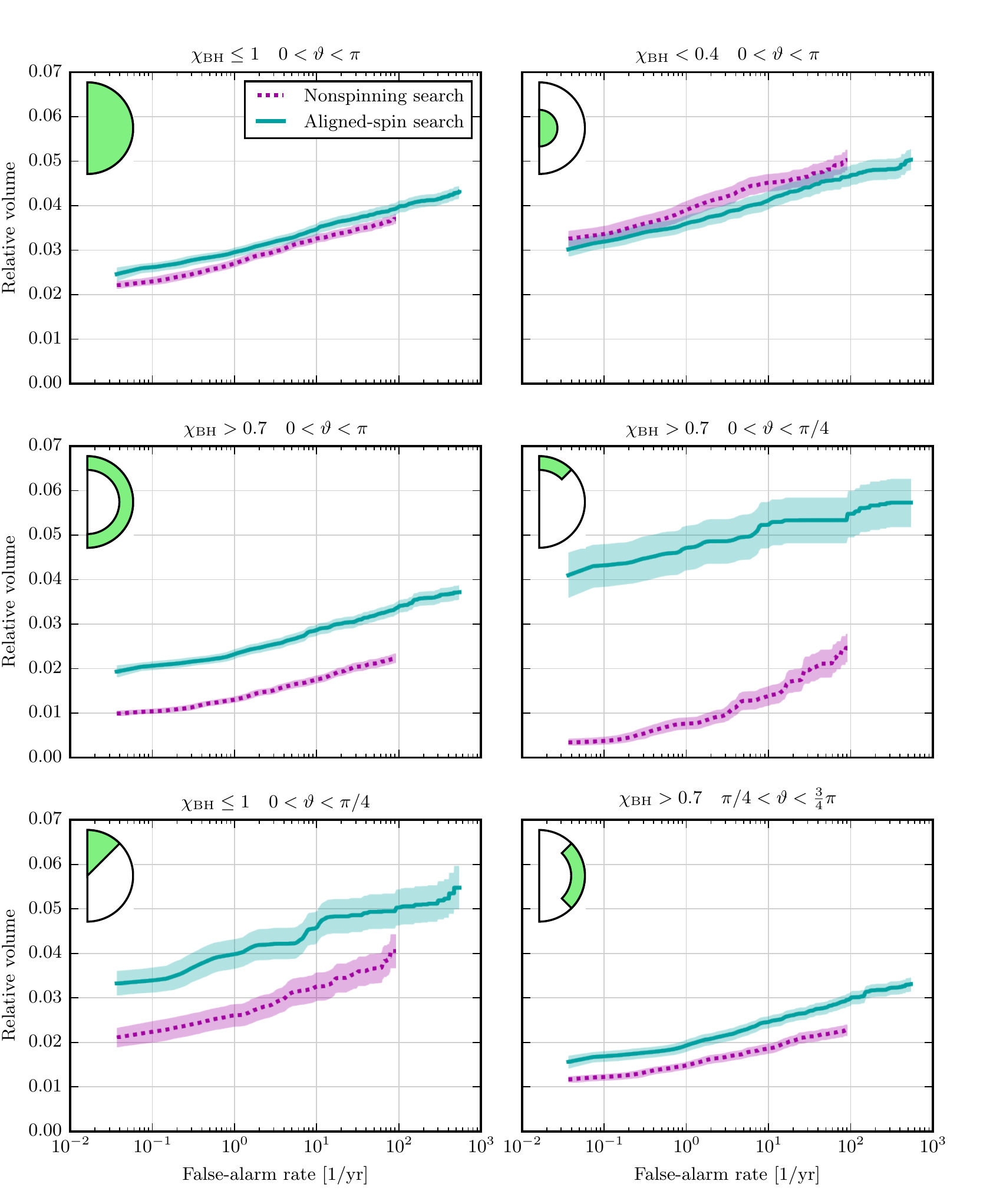}
    \caption{\acp{ROC} associated with nonspinning and spinning search pipelines
    (dashed and solid lines respectively) observing populations of \ac{NSBH}
    binaries with mass distribution as in Section \ref{sec:effectualness} but
    different constraints on the \ac{BH} spin parameters $(\chiperp,\chipar)$
    (visualized in the insets). The lighter bands show the 68\% error intervals
    estimated by constructing each curve 100 times from different combinations
    of the available data. Compare with the final plot of \cite{pycbcpaper}.}
    \label{fig:rocvarmass}
\end{figure*}

\acp{ROC} for the varying-mass population are shown in Figure
\ref{fig:rocvarmass}. The curves are qualitatively similar to the fixed-mass
case and we can thus extend our previous conclusions to a realistic distribution
of masses. The curves should be compared to the final plot of \cite{pycbcpaper}.

Similarly to what was done in Section \ref{sec:effectualness}, \acp{ROC} allow us
to estimate the detection rate of an aligned-spin search pipeline relative to
an ideal generic precessing pipeline, once we make the two following
approximations. The first is that the precessing pipeline produces the same
false-alarm background as the aligned-spin one. This is likely not the case, so
our result will overestimate the loss of detections due to neglecting
precession; nevertheless, based on the comparison between the false alarms of
the \ac{NSB} and \ac{ASB} in \cite{pycbcpaper}, we expect the error to be
small.  The second assumption is that the number of detections of the
precessing pipeline, averaged over the orientation of $\vec{J}$, is independent
from $\chiperp$; in other words, we assume that precessing sources are not
intrinsically more visible than non-precessing ones, when all orientations of
$\vec{J}$ are considered. This assumption is reasonable because precession
distributes the radiated power more evenly across the sky, but it does not make
the signal significantly longer or louder \cite{BLO}.  If these assumptions are
met, the fraction of generic sources detected by the precessing pipeline should
equal the fraction of weakly-precessing sources detected by the \ac{ASB}. Since
the fraction of detected sources is our definition of \ac{ROC}, the \ac{ROC} of
the \ac{ASB} for weakly-precessing sources (say, $\chiperp < 0.4$) can then be
used as a proxy for the precessing pipeline over the full spin parameter space.
We can thus estimate the relative sensitivity of the aligned-spin pipeline with
respect to a precessing one as
\begin{equation}
    \mathcal{W}^{\{\rm all,HP\}}_{\{\rm FM,VM\}}(f) \equiv
        \frac{W^{\{\rm all,HP\}}_{\{\rm FM,VM\}}(f)}{W^{\rm LP}_{\{\rm FM,VM\}}(f)}
\end{equation}
where $W^S_M(f)$ is the \ac{ROC} associated with the aligned-spin pipeline and
the particular cut $S$ of the spin parameter space (``all'': no restriction on
spin; ``LP'': low precession, $\chiperp < 0.4$; ``HP'': high precession,
$\chibh > 0.7$ and $\frac{1}{4}\pi < \vartheta < \frac{3}{4}\pi$) and mass
distribution $M$ (``FM'': fixed mass; ``VM'': varying mass) and $f$ is the
false-alarm rate at which the \ac{ROC} is evaluated. We obtain
\begin{eqnarray}
    78\% < &\mathcal{W}^{\rm all}_{\rm FM}& < 83\% \\
    41\% < &\mathcal{W}^{\rm HP}_{\rm FM}& < 50\% \\
    80\% < &\mathcal{W}^{\rm all}_{\rm VM}& < 87\% \\
    50\% < &\mathcal{W}^{\rm HP}_{\rm VM}& < 68\%
\end{eqnarray}
where the ranges include the different possibe false-alarm rates. As opposed to
the estimates obtained at the end of Section \ref{sec:effectualness}, these
estimates do include the effect of the $\chi^2$ veto. This seems to have a
small effect in the unrestricted-spin case, as $\mathcal{W}^{\rm all}_{\rm FM}$
and $\mathcal{W}^{\rm all}_{\rm VM}$ are consistent with $\mathcal{V}^{\rm
all}_{\rm FM}$ and $\mathcal{V}^{\rm all}_{\rm VM}$ respectively. However, the
high-precession estimates are systematically smaller than Eq.~\ref{eq:vhpfm}
and \ref{eq:vhpvm}, suggesting that the $\chi^2$ veto is indeed penalizing
precessing signals.

Relaxing the assumption of equal background of the precessing and aligned-spin
pipelines requires at least an estimate of the number of precessing templates,
but it can only increase all $\mathcal{W}$ estimates.  We therefore expect that
the loss of detections due to using an aligned-spin pipeline versus a
precessing one is, with no restrictions on \ac{BH} spin parameters, no larger
than $\sim 20\%$; if we make a less plausible assumption of highly-precessing
\ac{NSBH} binaries, we expect a loss within $\sim 60\%$.

\section{Conclusion}
\label{sec:conclusion}

In this paper we study the effect of precession on \ac{NSBH} binary inspiral
search pipelines using nonspinning and aligned-spin template banks.

By means of bank simulations, we first show that the two banks perform
similarly over the parameter space of the dimensionless \ac{BH} spin, except
for a large drop in effectualness of the nonspinning bank when the projection
of the spin on the orbital angular momentum is larger than $\sim 0.4$ or
smaller than $\sim -0.5$, which corresponds to strongly-spinning but
weakly-precessing systems. When precession is strong, both banks can lose up to
$\sim 40\%$ of the \ac{SNR} for particular orientations of the total angular
momentum with respect to the detector.  Nevertheless, the high-precession
systems which are best recovered correspond to face-on and face-off
orientations, which are the most likely to be observed based on their intrinsic
\ac{SNR}.

Using the same template banks, we then employ a realistic search pipeline to
recover the same signal population in simulated noise, constructed by
recoloring real initial-LIGO data to a sensitivity indicative of early
advanced-LIGO detectors.  The search pipeline includes the $\chi^2$
signal-based veto and exact-match coincidence between the Hanford and
Livingston LIGO detectors. We compare the resulting nonspinning and
aligned-spin \ac{ROC} curves for different choices of \ac{BH} spin parameters.

We conclude that using an aligned-spin bank will increase the detection rate of
\ac{NSBH} binaries by a fraction which strongly depends on the distribution of
\ac{BH} spin magnitudes and tilt angles in nature. At the very minimum,
assuming an extreme case of weakly-spinning \acp{BH} ($0 \leq \chibh < 0.4$)
and unrestricted, uniformly-distributed tilt angle, a search pipeline based on
the aligned-spin bank would have a larger computational cost but it would
reduce the sensitivity relative to using a nonspinning bank by $\sim 10\%$ at
most.  With unrestricted, uniformly-distributed spin magnitude and tilt, both
methods have very similar sensitivity.  On the other hand, for large spin
magnitude and small tilt---a distribution supported by population-synthesis
models and existing \ac{BH} spin measurements---the improvement in sensitivity
can be as large as one order of magnitude. We also find a noticeable
improvement assuming either small tilt and unrestricted magnitude, or
unrestricted tilt and large magnitude.  A factor of $\sim 2$ in sensitivity is
lost for the unlikely case of strongly-precessing systems, even with the
aligned-spin bank. Using these results we estimate that using a bank of generic
precessing templates could increase the sensitivity by tens of percent or,
under less realistic assumptions, possibly by a factor of 2, assuming the
background does not increase significantly. Quantifying this improvement more
precisely requires constructing a bank of precessing waveforms (see
e.g.~\cite{Buonanno:2002fy, PhysRevD.69.104017, Harry:2011qh, Lundgren:2013jla,
PhysRevD.89.084006, PhysRevLett.113.151101}) and applying it to a full search
pipeline with realistic advanced-detector data, which will be subjects of
future papers.

Although we assumed a particular sensitivity curve, our conclusion is likely to
remain true for the wider-band sensitivity expected for the final design
advanced LIGO. Bank simulations in \cite{pycbcpaper} suggest that including
spin is even more important for the zero-detuned high-power design curve of
advanced LIGO; similar calculations with precessing \ac{NSBH} signals also
produced comparable results \cite{Harry:2013tca, PhysRevD.89.084041}. The
simulated precessing signals we have used are based on post-Newtonian
expansions and terminate abruptly, ignoring merger and ringdown effects. For
the mass range considered here, except for highly anti-aligned systems, these
are likely to be good approximations to the true waveforms in the sensitive
band of advanced LIGO detectors; thus we do not expect the inclusion of merger
and ringdown to change our results significantly. Should the search be extended
to much higher masses---such that merger and ringdown provide the majority of
the \ac{SNR}---the inclusion of spin could be less important. Tidal interaction
between the \ac{NS} and \ac{BH} may be important in some regions of our
parameter space, for instance by altering the post-Newtonian phasing or
disrupting the \ac{NS} and shutting down the \ac{GW} signal before merger
\cite{PhysRevD.86.124007}. Based on \cite{Pannarale:2011pk}, we do not expect
tidal deformation to significantly affect the search. Tidal disruption should
also not affect our conclusions if it happens beyond the \ac{ISCO} frequency
used for our templates. We reserve studying the inclusion of merger and
ringdown, tidal effects and comparison with numerical waveforms to a future
paper.

Assuming the larger computational cost can be satisfied, we conclude with the
recommendation of switching from nonspinning to aligned-spin templates for
future \ac{NSBH} searches in the mass range considered here, as this would
provide good sensitivity across the entire \ac{BH} spin parameter space,
including both the aligned and precessing high-spin cases.

\begin{acknowledgments}
We thank Steve Privitera, Tom Dent and Badri Krishnan for useful discussion and
comments on the paper.  We also acknowledge the LIGO collaboration for
providing the recolored synthetic strain data we analyzed. TDC is supported by
the International Max-Planck Research School on Gravitational-Wave Astronomy.

This paper has LIGO document number P1400209.
\end{acknowledgments}

\bibliography{paper}{}

\begin{thebibliography}{38}%
\makeatletter
\providecommand \@ifxundefined [1]{%
 \@ifx{#1\undefined}
}%
\providecommand \@ifnum [1]{%
 \ifnum #1\expandafter \@firstoftwo
 \else \expandafter \@secondoftwo
 \fi
}%
\providecommand \@ifx [1]{%
 \ifx #1\expandafter \@firstoftwo
 \else \expandafter \@secondoftwo
 \fi
}%
\providecommand \natexlab [1]{#1}%
\providecommand \enquote  [1]{``#1''}%
\providecommand \bibnamefont  [1]{#1}%
\providecommand \bibfnamefont [1]{#1}%
\providecommand \citenamefont [1]{#1}%
\providecommand \href@noop [0]{\@secondoftwo}%
\providecommand \href [0]{\begingroup \@sanitize@url \@href}%
\providecommand \@href[1]{\@@startlink{#1}\@@href}%
\providecommand \@@href[1]{\endgroup#1\@@endlink}%
\providecommand \@sanitize@url [0]{\catcode `\\12\catcode `\$12\catcode
  `\&12\catcode `\#12\catcode `\^12\catcode `\_12\catcode `\%12\relax}%
\providecommand \@@startlink[1]{}%
\providecommand \@@endlink[0]{}%
\providecommand \url  [0]{\begingroup\@sanitize@url \@url }%
\providecommand \@url [1]{\endgroup\@href {#1}{\urlprefix }}%
\providecommand \urlprefix  [0]{URL }%
\providecommand \Eprint [0]{\href }%
\providecommand \doibase [0]{http://dx.doi.org/}%
\providecommand \selectlanguage [0]{\@gobble}%
\providecommand \bibinfo  [0]{\@secondoftwo}%
\providecommand \bibfield  [0]{\@secondoftwo}%
\providecommand \translation [1]{[#1]}%
\providecommand \BibitemOpen [0]{}%
\providecommand \bibitemStop [0]{}%
\providecommand \bibitemNoStop [0]{.\EOS\space}%
\providecommand \EOS [0]{\spacefactor3000\relax}%
\providecommand \BibitemShut  [1]{\csname bibitem#1\endcsname}%
\let\auto@bib@innerbib\@empty
\bibitem [{\citenamefont {Harry}(2010)}]{Harry:2010zz}%
  \BibitemOpen
  \bibfield  {author} {\bibinfo {author} {\bibfnamefont {G.~M.}\ \bibnamefont
  {Harry}} (\bibinfo {collaboration} {LIGO Scientific Collaboration}),\ }\href
  {\doibase 10.1088/0264-9381/27/8/084006} {\bibfield  {journal} {\bibinfo
  {journal} {Class.Quant.Grav.}\ }\textbf {\bibinfo {volume} {27}},\ \bibinfo
  {pages} {084006} (\bibinfo {year} {2010})}\BibitemShut {NoStop}%
\bibitem [{\citenamefont {Acernese}\ \emph {et~al.}(2015)\citenamefont
  {Acernese} \emph {et~al.}}]{TheVirgo:2014hva}%
  \BibitemOpen
  \bibfield  {author} {\bibinfo {author} {\bibfnamefont {F.}~\bibnamefont
  {Acernese}} \emph {et~al.} (\bibinfo {collaboration} {The Virgo
  Collaboration}),\ }\href {http://stacks.iop.org/0264-9381/32/i=2/a=024001}
  {\bibfield  {journal} {\bibinfo  {journal} {Classical and Quantum Gravity}\
  }\textbf {\bibinfo {volume} {32}},\ \bibinfo {pages} {024001} (\bibinfo
  {year} {2015})},\ \Eprint {http://arxiv.org/abs/1408.3978} {arXiv:1408.3978
  [gr-qc]} \BibitemShut {NoStop}%
\bibitem [{\citenamefont {Somiya}(2012)}]{Somiya:2011np}%
  \BibitemOpen
  \bibfield  {author} {\bibinfo {author} {\bibfnamefont {K.}~\bibnamefont
  {Somiya}} (\bibinfo {collaboration} {KAGRA Collaboration}),\ }\href {\doibase
  10.1088/0264-9381/29/12/124007} {\bibfield  {journal} {\bibinfo  {journal}
  {Class.Quant.Grav.}\ }\textbf {\bibinfo {volume} {29}},\ \bibinfo {pages}
  {124007} (\bibinfo {year} {2012})},\ \Eprint {http://arxiv.org/abs/1111.7185}
  {arXiv:1111.7185 [gr-qc]} \BibitemShut {NoStop}%
\bibitem [{\citenamefont {Abbott}\ \emph {et~al.}(2009)\citenamefont {Abbott}
  \emph {et~al.}}]{Abbott:2009tt}%
  \BibitemOpen
  \bibfield  {author} {\bibinfo {author} {\bibfnamefont {B.}~\bibnamefont
  {Abbott}} \emph {et~al.} (\bibinfo {collaboration} {LIGO Scientific
  Collaboration}),\ }\href {\doibase 10.1103/PhysRevD.79.122001} {\bibfield
  {journal} {\bibinfo  {journal} {Phys. Rev. D}\ }\textbf {\bibinfo {volume}
  {79}},\ \bibinfo {pages} {122001} (\bibinfo {year} {2009})},\ \Eprint
  {http://arxiv.org/abs/0901.0302} {arXiv:0901.0302 [gr-qc]} \BibitemShut
  {NoStop}%
\bibitem [{\citenamefont {Abadie}\ \emph {et~al.}(2010)\citenamefont {Abadie}
  \emph {et~al.}}]{Abadie:2010yb}%
  \BibitemOpen
  \bibfield  {author} {\bibinfo {author} {\bibfnamefont {J.}~\bibnamefont
  {Abadie}} \emph {et~al.} (\bibinfo {collaboration} {LIGO Scientific
  Collaboration, Virgo Collaboration}),\ }\href {\doibase
  10.1103/PhysRevD.85.089903, 10.1103/PhysRevD.82.102001} {\bibfield  {journal}
  {\bibinfo  {journal} {Phys. Rev. D}\ }\textbf {\bibinfo {volume} {82}},\
  \bibinfo {pages} {102001} (\bibinfo {year} {2010})},\ \Eprint
  {http://arxiv.org/abs/1005.4655} {arXiv:1005.4655 [gr-qc]} \BibitemShut
  {NoStop}%
\bibitem [{\citenamefont {Abadie}\ \emph {et~al.}(2011)\citenamefont {Abadie}
  \emph {et~al.}}]{Abadie:2011kd}%
  \BibitemOpen
  \bibfield  {author} {\bibinfo {author} {\bibfnamefont {J.}~\bibnamefont
  {Abadie}} \emph {et~al.} (\bibinfo {collaboration} {LIGO Scientific
  Collaboration, Virgo Collaboration}),\ }\href {\doibase
  10.1103/PhysRevD.86.069903, 10.1103/PhysRevD.85.089904,
  10.1103/PhysRevD.83.122005} {\bibfield  {journal} {\bibinfo  {journal} {Phys.
  Rev. D}\ }\textbf {\bibinfo {volume} {83}},\ \bibinfo {pages} {122005}
  (\bibinfo {year} {2011})},\ \Eprint {http://arxiv.org/abs/1102.3781}
  {arXiv:1102.3781 [gr-qc]} \BibitemShut {NoStop}%
\bibitem [{\citenamefont {Abadie}\ \emph {et~al.}(2012)\citenamefont {Abadie}
  \emph {et~al.}}]{Colaboration:2011np}%
  \BibitemOpen
  \bibfield  {author} {\bibinfo {author} {\bibfnamefont {J.}~\bibnamefont
  {Abadie}} \emph {et~al.} (\bibinfo {collaboration} {LIGO Collaboration, Virgo
  Collaboration}),\ }\href {\doibase 10.1103/PhysRevD.85.082002} {\bibfield
  {journal} {\bibinfo  {journal} {Phys. Rev. D}\ }\textbf {\bibinfo {volume}
  {85}},\ \bibinfo {pages} {082002} (\bibinfo {year} {2012})},\ \Eprint
  {http://arxiv.org/abs/1111.7314} {arXiv:1111.7314 [gr-qc]} \BibitemShut
  {NoStop}%
\bibitem [{\citenamefont {Aasi}\ \emph {et~al.}(2013)\citenamefont {Aasi} \emph
  {et~al.}}]{oai:arXiv.org:1209.6533}%
  \BibitemOpen
  \bibfield  {author} {\bibinfo {author} {\bibfnamefont {J.}~\bibnamefont
  {Aasi}} \emph {et~al.} (\bibinfo {collaboration} {LIGO Scientific
  Collaboration, Virgo Collaboration}),\ }\href {\doibase
  10.1103/PhysRevD.87.022002} {\bibfield  {journal} {\bibinfo  {journal} {Phys.
  Rev. D}\ }\textbf {\bibinfo {volume} {87}},\ \bibinfo {pages} {022002}
  (\bibinfo {year} {2013})},\ \Eprint {http://arxiv.org/abs/1209.6533}
  {arXiv:1209.6533 [gr-qc]} \BibitemShut {NoStop}%
\bibitem [{\citenamefont {Brown}\ \emph
  {et~al.}(2012{\natexlab{a}})\citenamefont {Brown}, \citenamefont {Harry},
  \citenamefont {Lundgren},\ and\ \citenamefont {Nitz}}]{Brown:2012qf}%
  \BibitemOpen
  \bibfield  {author} {\bibinfo {author} {\bibfnamefont {D.~A.}\ \bibnamefont
  {Brown}}, \bibinfo {author} {\bibfnamefont {I.}~\bibnamefont {Harry}},
  \bibinfo {author} {\bibfnamefont {A.}~\bibnamefont {Lundgren}}, \ and\
  \bibinfo {author} {\bibfnamefont {A.~H.}\ \bibnamefont {Nitz}},\ }\href
  {\doibase 10.1103/PhysRevD.86.084017} {\bibfield  {journal} {\bibinfo
  {journal} {Phys. Rev. D}\ }\textbf {\bibinfo {volume} {86}},\ \bibinfo
  {pages} {084017} (\bibinfo {year} {2012}{\natexlab{a}})},\ \Eprint
  {http://arxiv.org/abs/1207.6406} {arXiv:1207.6406 [gr-qc]} \BibitemShut
  {NoStop}%
\bibitem [{\citenamefont {Ajith}\ \emph {et~al.}(2014)\citenamefont {Ajith},
  \citenamefont {Fotopoulos}, \citenamefont {Privitera}, \citenamefont
  {Neunzert}, \citenamefont {Mazumder},\ and\ \citenamefont
  {Weinstein}}]{PhysRevD.89.084041}%
  \BibitemOpen
  \bibfield  {author} {\bibinfo {author} {\bibfnamefont {P.}~\bibnamefont
  {Ajith}}, \bibinfo {author} {\bibfnamefont {N.}~\bibnamefont {Fotopoulos}},
  \bibinfo {author} {\bibfnamefont {S.}~\bibnamefont {Privitera}}, \bibinfo
  {author} {\bibfnamefont {A.}~\bibnamefont {Neunzert}}, \bibinfo {author}
  {\bibfnamefont {N.}~\bibnamefont {Mazumder}}, \ and\ \bibinfo {author}
  {\bibfnamefont {A.~J.}\ \bibnamefont {Weinstein}},\ }\href {\doibase
  10.1103/PhysRevD.89.084041} {\bibfield  {journal} {\bibinfo  {journal} {Phys.
  Rev. D}\ }\textbf {\bibinfo {volume} {89}},\ \bibinfo {pages} {084041}
  (\bibinfo {year} {2014})},\ \Eprint {http://arxiv.org/abs/1210.6666}
  {arXiv:1210.6666 [gr-qc]} \BibitemShut {NoStop}%
\bibitem [{\citenamefont {Dal~Canton}\ \emph {et~al.}(2014)\citenamefont
  {Dal~Canton}, \citenamefont {Nitz}, \citenamefont {Lundgren}, \citenamefont
  {Nielsen}, \citenamefont {Brown}, \citenamefont {Dent}, \citenamefont
  {Harry}, \citenamefont {Krishnan}, \citenamefont {Miller}, \citenamefont
  {Wette}, \citenamefont {Wiesner},\ and\ \citenamefont {Willis}}]{pycbcpaper}%
  \BibitemOpen
  \bibfield  {author} {\bibinfo {author} {\bibfnamefont {T.}~\bibnamefont
  {Dal~Canton}}, \bibinfo {author} {\bibfnamefont {A.~H.}\ \bibnamefont
  {Nitz}}, \bibinfo {author} {\bibfnamefont {A.~P.}\ \bibnamefont {Lundgren}},
  \bibinfo {author} {\bibfnamefont {A.~B.}\ \bibnamefont {Nielsen}}, \bibinfo
  {author} {\bibfnamefont {D.~A.}\ \bibnamefont {Brown}}, \bibinfo {author}
  {\bibfnamefont {T.}~\bibnamefont {Dent}}, \bibinfo {author} {\bibfnamefont
  {I.~W.}\ \bibnamefont {Harry}}, \bibinfo {author} {\bibfnamefont
  {B.}~\bibnamefont {Krishnan}}, \bibinfo {author} {\bibfnamefont {A.~J.}\
  \bibnamefont {Miller}}, \bibinfo {author} {\bibfnamefont {K.}~\bibnamefont
  {Wette}}, \bibinfo {author} {\bibfnamefont {K.}~\bibnamefont {Wiesner}}, \
  and\ \bibinfo {author} {\bibfnamefont {J.~L.}\ \bibnamefont {Willis}},\
  }\href {\doibase 10.1103/PhysRevD.90.082004} {\bibfield  {journal} {\bibinfo
  {journal} {Phys. Rev. D}\ }\textbf {\bibinfo {volume} {90}},\ \bibinfo
  {pages} {082004} (\bibinfo {year} {2014})},\ \Eprint
  {http://arxiv.org/abs/1405.6731} {arXiv:1405.6731 [gr-qc]} \BibitemShut
  {NoStop}%
\bibitem [{\citenamefont {O'Shaughnessy}\ \emph {et~al.}(2005)\citenamefont
  {O'Shaughnessy}, \citenamefont {Kaplan}, \citenamefont {Kalogera},\ and\
  \citenamefont {Belczynski}}]{O'Shaughnessy:2005qc}%
  \BibitemOpen
  \bibfield  {author} {\bibinfo {author} {\bibfnamefont {R.~W.}\ \bibnamefont
  {O'Shaughnessy}}, \bibinfo {author} {\bibfnamefont {J.}~\bibnamefont
  {Kaplan}}, \bibinfo {author} {\bibfnamefont {V.}~\bibnamefont {Kalogera}}, \
  and\ \bibinfo {author} {\bibfnamefont {K.}~\bibnamefont {Belczynski}},\
  }\href {\doibase 10.1086/444346} {\bibfield  {journal} {\bibinfo  {journal}
  {Astrophys.J.}\ }\textbf {\bibinfo {volume} {632}},\ \bibinfo {pages} {1035}
  (\bibinfo {year} {2005})},\ \Eprint {http://arxiv.org/abs/astro-ph/0503219}
  {arXiv:astro-ph/0503219 [astro-ph]} \BibitemShut {NoStop}%
\bibitem [{\citenamefont {Dominik}\ \emph {et~al.}(2012)\citenamefont
  {Dominik}, \citenamefont {Belczynski}, \citenamefont {Fryer}, \citenamefont
  {Holz}, \citenamefont {Berti} \emph {et~al.}}]{Dominik:2012kk}%
  \BibitemOpen
  \bibfield  {author} {\bibinfo {author} {\bibfnamefont {M.}~\bibnamefont
  {Dominik}}, \bibinfo {author} {\bibfnamefont {K.}~\bibnamefont {Belczynski}},
  \bibinfo {author} {\bibfnamefont {C.}~\bibnamefont {Fryer}}, \bibinfo
  {author} {\bibfnamefont {D.}~\bibnamefont {Holz}}, \bibinfo {author}
  {\bibfnamefont {E.}~\bibnamefont {Berti}},  \emph {et~al.},\ }\href {\doibase
  10.1088/0004-637X/759/1/52} {\bibfield  {journal} {\bibinfo  {journal}
  {Astrophys.J.}\ }\textbf {\bibinfo {volume} {759}},\ \bibinfo {pages} {52}
  (\bibinfo {year} {2012})},\ \Eprint {http://arxiv.org/abs/1202.4901}
  {arXiv:1202.4901 [astro-ph.HE]} \BibitemShut {NoStop}%
\bibitem [{\citenamefont {Kalogera}(2000)}]{Kalogera:1999tq}%
  \BibitemOpen
  \bibfield  {author} {\bibinfo {author} {\bibfnamefont {V.}~\bibnamefont
  {Kalogera}},\ }\href {\doibase 10.1086/309400} {\bibfield  {journal}
  {\bibinfo  {journal} {Astrophys.J.}\ }\textbf {\bibinfo {volume} {541}},\
  \bibinfo {pages} {319} (\bibinfo {year} {2000})},\ \Eprint
  {http://arxiv.org/abs/astro-ph/9911417} {arXiv:astro-ph/9911417 [astro-ph]}
  \BibitemShut {NoStop}%
\bibitem [{\citenamefont {Grandcl\'ement}\ \emph {et~al.}(2004)\citenamefont
  {Grandcl\'ement}, \citenamefont {Ihm}, \citenamefont {Kalogera},\ and\
  \citenamefont {Belczynski}}]{PhysRevD.69.102002}%
  \BibitemOpen
  \bibfield  {author} {\bibinfo {author} {\bibfnamefont {P.}~\bibnamefont
  {Grandcl\'ement}}, \bibinfo {author} {\bibfnamefont {M.}~\bibnamefont {Ihm}},
  \bibinfo {author} {\bibfnamefont {V.}~\bibnamefont {Kalogera}}, \ and\
  \bibinfo {author} {\bibfnamefont {K.}~\bibnamefont {Belczynski}},\ }\href
  {\doibase 10.1103/PhysRevD.69.102002} {\bibfield  {journal} {\bibinfo
  {journal} {Phys. Rev. D}\ }\textbf {\bibinfo {volume} {69}},\ \bibinfo
  {pages} {102002} (\bibinfo {year} {2004})},\ \Eprint
  {http://arxiv.org/abs/gr-qc/0312084} {arXiv:gr-qc/0312084 [gr-qc]}
  \BibitemShut {NoStop}%
\bibitem [{\citenamefont {Belczynski}\ \emph {et~al.}(2008)\citenamefont
  {Belczynski}, \citenamefont {Taam}, \citenamefont {Rantsiou},\ and\
  \citenamefont {van~der Sluys}}]{Belczynski:2007xg}%
  \BibitemOpen
  \bibfield  {author} {\bibinfo {author} {\bibfnamefont {K.}~\bibnamefont
  {Belczynski}}, \bibinfo {author} {\bibfnamefont {R.~E.}\ \bibnamefont
  {Taam}}, \bibinfo {author} {\bibfnamefont {E.}~\bibnamefont {Rantsiou}}, \
  and\ \bibinfo {author} {\bibfnamefont {M.}~\bibnamefont {van~der Sluys}},\
  }\href {http://stacks.iop.org/0004-637X/682/i=1/a=474} {\bibfield  {journal}
  {\bibinfo  {journal} {The Astrophysical Journal}\ }\textbf {\bibinfo {volume}
  {682}},\ \bibinfo {pages} {474} (\bibinfo {year} {2008})},\ \Eprint
  {http://arxiv.org/abs/astro-ph/0703131} {arXiv:astro-ph/0703131 [ASTRO-PH]}
  \BibitemShut {NoStop}%
\bibitem [{\citenamefont {Fragos}\ \emph {et~al.}(2010)\citenamefont {Fragos},
  \citenamefont {Tremmel}, \citenamefont {Rantsiou},\ and\ \citenamefont
  {Belczynski}}]{Fragos:2010tm}%
  \BibitemOpen
  \bibfield  {author} {\bibinfo {author} {\bibfnamefont {T.}~\bibnamefont
  {Fragos}}, \bibinfo {author} {\bibfnamefont {M.}~\bibnamefont {Tremmel}},
  \bibinfo {author} {\bibfnamefont {E.}~\bibnamefont {Rantsiou}}, \ and\
  \bibinfo {author} {\bibfnamefont {K.}~\bibnamefont {Belczynski}},\ }\href
  {http://stacks.iop.org/2041-8205/719/i=1/a=L79} {\bibfield  {journal}
  {\bibinfo  {journal} {The Astrophysical Journal Letters}\ }\textbf {\bibinfo
  {volume} {719}},\ \bibinfo {pages} {L79} (\bibinfo {year} {2010})},\ \Eprint
  {http://arxiv.org/abs/1001.1107} {arXiv:1001.1107 [astro-ph.HE]} \BibitemShut
  {NoStop}%
\bibitem [{\citenamefont {Fragile}(2009)}]{Fragile:2009vu}%
  \BibitemOpen
  \bibfield  {author} {\bibinfo {author} {\bibfnamefont {P.~C.}\ \bibnamefont
  {Fragile}},\ }\href {\doibase 10.1088/0004-637X/706/2/L246} {\bibfield
  {journal} {\bibinfo  {journal} {Astrophys.J.}\ }\textbf {\bibinfo {volume}
  {706}},\ \bibinfo {pages} {L246} (\bibinfo {year} {2009})},\ \Eprint
  {http://arxiv.org/abs/0910.5721} {arXiv:0910.5721 [astro-ph.HE]} \BibitemShut
  {NoStop}%
\bibitem [{\citenamefont {Apostolatos}\ \emph {et~al.}(1994)\citenamefont
  {Apostolatos}, \citenamefont {Cutler}, \citenamefont {Sussman},\ and\
  \citenamefont {Thorne}}]{Apostolatos:1994mx}%
  \BibitemOpen
  \bibfield  {author} {\bibinfo {author} {\bibfnamefont {T.~A.}\ \bibnamefont
  {Apostolatos}}, \bibinfo {author} {\bibfnamefont {C.}~\bibnamefont {Cutler}},
  \bibinfo {author} {\bibfnamefont {G.~J.}\ \bibnamefont {Sussman}}, \ and\
  \bibinfo {author} {\bibfnamefont {K.~S.}\ \bibnamefont {Thorne}},\ }\href
  {\doibase 10.1103/PhysRevD.49.6274} {\bibfield  {journal} {\bibinfo
  {journal} {Phys. Rev. D}\ }\textbf {\bibinfo {volume} {49}},\ \bibinfo
  {pages} {6274} (\bibinfo {year} {1994})}\BibitemShut {NoStop}%
\bibitem [{\citenamefont {Brown}\ \emph
  {et~al.}(2012{\natexlab{b}})\citenamefont {Brown}, \citenamefont {Lundgren},\
  and\ \citenamefont {O'Shaughnessy}}]{BLO}%
  \BibitemOpen
  \bibfield  {author} {\bibinfo {author} {\bibfnamefont {D.~A.}\ \bibnamefont
  {Brown}}, \bibinfo {author} {\bibfnamefont {A.}~\bibnamefont {Lundgren}}, \
  and\ \bibinfo {author} {\bibfnamefont {R.}~\bibnamefont {O'Shaughnessy}},\
  }\href {\doibase 10.1103/PhysRevD.86.064020} {\bibfield  {journal} {\bibinfo
  {journal} {Phys. Rev. D}\ }\textbf {\bibinfo {volume} {86}},\ \bibinfo
  {pages} {064020} (\bibinfo {year} {2012}{\natexlab{b}})},\ \Eprint
  {http://arxiv.org/abs/1203.6060} {arXiv:1203.6060 [gr-qc]} \BibitemShut
  {NoStop}%
\bibitem [{\citenamefont {Van Den~Broeck}\ \emph {et~al.}(2009)\citenamefont
  {Van Den~Broeck}, \citenamefont {Brown}, \citenamefont {Cokelaer},
  \citenamefont {Harry}, \citenamefont {Jones}, \citenamefont {Sathyaprakash},
  \citenamefont {Tagoshi},\ and\ \citenamefont
  {Takahashi}}]{VanDenBroeck:2009gd}%
  \BibitemOpen
  \bibfield  {author} {\bibinfo {author} {\bibfnamefont {C.}~\bibnamefont {Van
  Den~Broeck}}, \bibinfo {author} {\bibfnamefont {D.~A.}\ \bibnamefont
  {Brown}}, \bibinfo {author} {\bibfnamefont {T.}~\bibnamefont {Cokelaer}},
  \bibinfo {author} {\bibfnamefont {I.}~\bibnamefont {Harry}}, \bibinfo
  {author} {\bibfnamefont {G.}~\bibnamefont {Jones}}, \bibinfo {author}
  {\bibfnamefont {B.~S.}\ \bibnamefont {Sathyaprakash}}, \bibinfo {author}
  {\bibfnamefont {H.}~\bibnamefont {Tagoshi}}, \ and\ \bibinfo {author}
  {\bibfnamefont {H.}~\bibnamefont {Takahashi}},\ }\href {\doibase
  10.1103/PhysRevD.80.024009} {\bibfield  {journal} {\bibinfo  {journal} {Phys.
  Rev. D}\ }\textbf {\bibinfo {volume} {80}},\ \bibinfo {pages} {024009}
  (\bibinfo {year} {2009})},\ \Eprint {http://arxiv.org/abs/0904.1715}
  {arXiv:0904.1715 [gr-qc]} \BibitemShut {NoStop}%
\bibitem [{\citenamefont {Abbott}\ \emph {et~al.}(2008)\citenamefont {Abbott}
  \emph {et~al.}}]{Abbott:2007ai}%
  \BibitemOpen
  \bibfield  {author} {\bibinfo {author} {\bibfnamefont {B.}~\bibnamefont
  {Abbott}} \emph {et~al.} (\bibinfo {collaboration} {LIGO Scientific
  Collaboration}),\ }\href {\doibase 10.1103/PhysRevD.78.042002} {\bibfield
  {journal} {\bibinfo  {journal} {Phys. Rev. D}\ }\textbf {\bibinfo {volume}
  {78}},\ \bibinfo {pages} {042002} (\bibinfo {year} {2008})},\ \Eprint
  {http://arxiv.org/abs/0712.2050} {arXiv:0712.2050 [gr-qc]} \BibitemShut
  {NoStop}%
\bibitem [{\citenamefont {Harry}\ \emph {et~al.}(2014)\citenamefont {Harry},
  \citenamefont {Nitz}, \citenamefont {Brown}, \citenamefont {Lundgren},
  \citenamefont {Ochsner},\ and\ \citenamefont {Keppel}}]{Harry:2013tca}%
  \BibitemOpen
  \bibfield  {author} {\bibinfo {author} {\bibfnamefont {I.~W.}\ \bibnamefont
  {Harry}}, \bibinfo {author} {\bibfnamefont {A.~H.}\ \bibnamefont {Nitz}},
  \bibinfo {author} {\bibfnamefont {D.~A.}\ \bibnamefont {Brown}}, \bibinfo
  {author} {\bibfnamefont {A.~P.}\ \bibnamefont {Lundgren}}, \bibinfo {author}
  {\bibfnamefont {E.}~\bibnamefont {Ochsner}}, \ and\ \bibinfo {author}
  {\bibfnamefont {D.}~\bibnamefont {Keppel}},\ }\href {\doibase
  10.1103/PhysRevD.89.024010} {\bibfield  {journal} {\bibinfo  {journal} {Phys.
  Rev. D}\ }\textbf {\bibinfo {volume} {89}},\ \bibinfo {pages} {024010}
  (\bibinfo {year} {2014})},\ \Eprint {http://arxiv.org/abs/1307.3562}
  {arXiv:1307.3562 [gr-qc]} \BibitemShut {NoStop}%
\bibitem [{\citenamefont {Barsotti}\ and\ \citenamefont
  {Fritschel}(2014)}]{LIGO-T1200307-v4}%
  \BibitemOpen
  \bibfield  {author} {\bibinfo {author} {\bibfnamefont {L.}~\bibnamefont
  {Barsotti}}\ and\ \bibinfo {author} {\bibfnamefont {P.}~\bibnamefont
  {Fritschel}},\ }\href
  {https://dcc.ligo.org/cgi-bin/DocDB/ShowDocument?.submit=Number&docid=T1200307&version=4}
  {\bibfield  {journal} {\bibinfo  {journal} {LIGO Technical Note}\ }\textbf
  {\bibinfo {volume} {T1200307-v4}} (\bibinfo {year} {2014})}\BibitemShut
  {NoStop}%
\bibitem [{\citenamefont {McClintock}\ \emph {et~al.}(2011)\citenamefont
  {McClintock}, \citenamefont {Narayan}, \citenamefont {Davis}, \citenamefont
  {Gou}, \citenamefont {Kulkarni} \emph {et~al.}}]{McClintock:2011zq}%
  \BibitemOpen
  \bibfield  {author} {\bibinfo {author} {\bibfnamefont {J.~E.}\ \bibnamefont
  {McClintock}}, \bibinfo {author} {\bibfnamefont {R.}~\bibnamefont {Narayan}},
  \bibinfo {author} {\bibfnamefont {S.~W.}\ \bibnamefont {Davis}}, \bibinfo
  {author} {\bibfnamefont {L.}~\bibnamefont {Gou}}, \bibinfo {author}
  {\bibfnamefont {A.}~\bibnamefont {Kulkarni}},  \emph {et~al.},\ }\href
  {\doibase 10.1088/0264-9381/28/11/114009} {\bibfield  {journal} {\bibinfo
  {journal} {Class.Quant.Grav.}\ }\textbf {\bibinfo {volume} {28}},\ \bibinfo
  {pages} {114009} (\bibinfo {year} {2011})},\ \Eprint
  {http://arxiv.org/abs/1101.0811} {arXiv:1101.0811 [astro-ph.HE]} \BibitemShut
  {NoStop}%
\bibitem [{\citenamefont {Damour}\ \emph {et~al.}(2012)\citenamefont {Damour},
  \citenamefont {Nagar},\ and\ \citenamefont {Villain}}]{Damour:2012yf}%
  \BibitemOpen
  \bibfield  {author} {\bibinfo {author} {\bibfnamefont {T.}~\bibnamefont
  {Damour}}, \bibinfo {author} {\bibfnamefont {A.}~\bibnamefont {Nagar}}, \
  and\ \bibinfo {author} {\bibfnamefont {L.}~\bibnamefont {Villain}},\ }\href
  {\doibase 10.1103/PhysRevD.85.123007} {\bibfield  {journal} {\bibinfo
  {journal} {Phys. Rev. D}\ }\textbf {\bibinfo {volume} {85}},\ \bibinfo
  {pages} {123007} (\bibinfo {year} {2012})},\ \Eprint
  {http://arxiv.org/abs/1203.4352} {arXiv:1203.4352 [gr-qc]} \BibitemShut
  {NoStop}%
\bibitem [{\citenamefont {\"Ozel}\ \emph {et~al.}(2010)\citenamefont {\"Ozel},
  \citenamefont {Psaltis}, \citenamefont {Narayan},\ and\ \citenamefont
  {McClintock}}]{Ozel:2010su}%
  \BibitemOpen
  \bibfield  {author} {\bibinfo {author} {\bibfnamefont {F.}~\bibnamefont
  {\"Ozel}}, \bibinfo {author} {\bibfnamefont {D.}~\bibnamefont {Psaltis}},
  \bibinfo {author} {\bibfnamefont {R.}~\bibnamefont {Narayan}}, \ and\
  \bibinfo {author} {\bibfnamefont {J.~E.}\ \bibnamefont {McClintock}},\ }\href
  {\doibase 10.1088/0004-637X/725/2/1918} {\bibfield  {journal} {\bibinfo
  {journal} {Astrophys.J.}\ }\textbf {\bibinfo {volume} {725}},\ \bibinfo
  {pages} {1918} (\bibinfo {year} {2010})},\ \Eprint
  {http://arxiv.org/abs/1006.2834} {arXiv:1006.2834 [astro-ph.GA]} \BibitemShut
  {NoStop}%
\bibitem [{\citenamefont {\"Ozel}\ \emph {et~al.}(2012)\citenamefont {\"Ozel},
  \citenamefont {Psaltis}, \citenamefont {Narayan},\ and\ \citenamefont
  {{Santos Villarreal}}}]{Ozel:2012ax}%
  \BibitemOpen
  \bibfield  {author} {\bibinfo {author} {\bibfnamefont {F.}~\bibnamefont
  {\"Ozel}}, \bibinfo {author} {\bibfnamefont {D.}~\bibnamefont {Psaltis}},
  \bibinfo {author} {\bibfnamefont {R.}~\bibnamefont {Narayan}}, \ and\
  \bibinfo {author} {\bibfnamefont {A.}~\bibnamefont {{Santos Villarreal}}},\
  }\href {\doibase 10.1088/0004-637X/757/1/55} {\bibfield  {journal} {\bibinfo
  {journal} {Astrophys.J.}\ }\textbf {\bibinfo {volume} {757}},\ \bibinfo
  {pages} {55} (\bibinfo {year} {2012})},\ \Eprint
  {http://arxiv.org/abs/1201.1006} {arXiv:1201.1006 [astro-ph.HE]} \BibitemShut
  {NoStop}%
\bibitem [{\citenamefont {Baird}\ \emph {et~al.}(2013)\citenamefont {Baird},
  \citenamefont {Fairhurst}, \citenamefont {Hannam},\ and\ \citenamefont
  {Murphy}}]{PhysRevD.87.024035}%
  \BibitemOpen
  \bibfield  {author} {\bibinfo {author} {\bibfnamefont {E.}~\bibnamefont
  {Baird}}, \bibinfo {author} {\bibfnamefont {S.}~\bibnamefont {Fairhurst}},
  \bibinfo {author} {\bibfnamefont {M.}~\bibnamefont {Hannam}}, \ and\ \bibinfo
  {author} {\bibfnamefont {P.}~\bibnamefont {Murphy}},\ }\href {\doibase
  10.1103/PhysRevD.87.024035} {\bibfield  {journal} {\bibinfo  {journal} {Phys.
  Rev. D}\ }\textbf {\bibinfo {volume} {87}},\ \bibinfo {pages} {024035}
  (\bibinfo {year} {2013})}\BibitemShut {NoStop}%
\bibitem [{\citenamefont {Lundgren}\ and\ \citenamefont
  {O'Shaughnessy}(2014)}]{Lundgren:2013jla}%
  \BibitemOpen
  \bibfield  {author} {\bibinfo {author} {\bibfnamefont {A.}~\bibnamefont
  {Lundgren}}\ and\ \bibinfo {author} {\bibfnamefont {R.}~\bibnamefont
  {O'Shaughnessy}},\ }\href {\doibase 10.1103/PhysRevD.89.044021} {\bibfield
  {journal} {\bibinfo  {journal} {Phys. Rev. D}\ }\textbf {\bibinfo {volume}
  {89}},\ \bibinfo {pages} {044021} (\bibinfo {year} {2014})},\ \Eprint
  {http://arxiv.org/abs/1304.3332} {arXiv:1304.3332 [gr-qc]} \BibitemShut
  {NoStop}%
\bibitem [{\citenamefont {Schmidt}\ \emph {et~al.}(2015)\citenamefont
  {Schmidt}, \citenamefont {Ohme},\ and\ \citenamefont
  {Hannam}}]{Schmidt:2014iyl}%
  \BibitemOpen
  \bibfield  {author} {\bibinfo {author} {\bibfnamefont {P.}~\bibnamefont
  {Schmidt}}, \bibinfo {author} {\bibfnamefont {F.}~\bibnamefont {Ohme}}, \
  and\ \bibinfo {author} {\bibfnamefont {M.}~\bibnamefont {Hannam}},\ }\href
  {\doibase 10.1103/PhysRevD.91.024043} {\bibfield  {journal} {\bibinfo
  {journal} {Phys. Rev. D}\ }\textbf {\bibinfo {volume} {91}},\ \bibinfo
  {pages} {024043} (\bibinfo {year} {2015})},\ \Eprint
  {http://arxiv.org/abs/1408.1810} {arXiv:1408.1810 [gr-qc]} \BibitemShut
  {NoStop}%
\bibitem [{\citenamefont {Buonanno}\ \emph {et~al.}(2003)\citenamefont
  {Buonanno}, \citenamefont {Chen},\ and\ \citenamefont
  {Vallisneri}}]{Buonanno:2002fy}%
  \BibitemOpen
  \bibfield  {author} {\bibinfo {author} {\bibfnamefont {A.}~\bibnamefont
  {Buonanno}}, \bibinfo {author} {\bibfnamefont {Y.-b.}\ \bibnamefont {Chen}},
  \ and\ \bibinfo {author} {\bibfnamefont {M.}~\bibnamefont {Vallisneri}},\
  }\href {\doibase 10.1103/PhysRevD.67.104025, 10.1103/PhysRevD.74.029904}
  {\bibfield  {journal} {\bibinfo  {journal} {Phys. Rev. D}\ }\textbf {\bibinfo
  {volume} {67}},\ \bibinfo {pages} {104025} (\bibinfo {year} {2003})},\
  \Eprint {http://arxiv.org/abs/gr-qc/0211087} {arXiv:gr-qc/0211087 [gr-qc]}
  \BibitemShut {NoStop}%
\bibitem [{\citenamefont {Pan}\ \emph {et~al.}(2004)\citenamefont {Pan},
  \citenamefont {Buonanno}, \citenamefont {Chen},\ and\ \citenamefont
  {Vallisneri}}]{PhysRevD.69.104017}%
  \BibitemOpen
  \bibfield  {author} {\bibinfo {author} {\bibfnamefont {Y.}~\bibnamefont
  {Pan}}, \bibinfo {author} {\bibfnamefont {A.}~\bibnamefont {Buonanno}},
  \bibinfo {author} {\bibfnamefont {Y.}~\bibnamefont {Chen}}, \ and\ \bibinfo
  {author} {\bibfnamefont {M.}~\bibnamefont {Vallisneri}},\ }\href {\doibase
  10.1103/PhysRevD.69.104017} {\bibfield  {journal} {\bibinfo  {journal} {Phys.
  Rev. D}\ }\textbf {\bibinfo {volume} {69}},\ \bibinfo {pages} {104017}
  (\bibinfo {year} {2004})},\ \Eprint {http://arxiv.org/abs/gr-qc/0310034}
  {arXiv:gr-qc/0310034 [gr-qc]} \BibitemShut {NoStop}%
\bibitem [{\citenamefont {Harry}\ and\ \citenamefont
  {Fairhurst}(2011)}]{Harry:2011qh}%
  \BibitemOpen
  \bibfield  {author} {\bibinfo {author} {\bibfnamefont {I.}~\bibnamefont
  {Harry}}\ and\ \bibinfo {author} {\bibfnamefont {S.}~\bibnamefont
  {Fairhurst}},\ }\href {\doibase 10.1088/0264-9381/28/13/134008} {\bibfield
  {journal} {\bibinfo  {journal} {Class.Quant.Grav.}\ }\textbf {\bibinfo
  {volume} {28}},\ \bibinfo {pages} {134008} (\bibinfo {year} {2011})},\
  \Eprint {http://arxiv.org/abs/1101.1459} {arXiv:1101.1459 [gr-qc]}
  \BibitemShut {NoStop}%
\bibitem [{\citenamefont {Pan}\ \emph {et~al.}(2014)\citenamefont {Pan},
  \citenamefont {Buonanno}, \citenamefont {Taracchini}, \citenamefont {Kidder},
  \citenamefont {Mrou\'e}, \citenamefont {Pfeiffer}, \citenamefont {Scheel},\
  and\ \citenamefont {Szil\'agyi}}]{PhysRevD.89.084006}%
  \BibitemOpen
  \bibfield  {author} {\bibinfo {author} {\bibfnamefont {Y.}~\bibnamefont
  {Pan}}, \bibinfo {author} {\bibfnamefont {A.}~\bibnamefont {Buonanno}},
  \bibinfo {author} {\bibfnamefont {A.}~\bibnamefont {Taracchini}}, \bibinfo
  {author} {\bibfnamefont {L.~E.}\ \bibnamefont {Kidder}}, \bibinfo {author}
  {\bibfnamefont {A.~H.}\ \bibnamefont {Mrou\'e}}, \bibinfo {author}
  {\bibfnamefont {H.~P.}\ \bibnamefont {Pfeiffer}}, \bibinfo {author}
  {\bibfnamefont {M.~A.}\ \bibnamefont {Scheel}}, \ and\ \bibinfo {author}
  {\bibfnamefont {B.}~\bibnamefont {Szil\'agyi}},\ }\href {\doibase
  10.1103/PhysRevD.89.084006} {\bibfield  {journal} {\bibinfo  {journal} {Phys.
  Rev. D}\ }\textbf {\bibinfo {volume} {89}},\ \bibinfo {pages} {084006}
  (\bibinfo {year} {2014})},\ \Eprint {http://arxiv.org/abs/1307.6232}
  {arXiv:1307.6232 [gr-qc]} \BibitemShut {NoStop}%
\bibitem [{\citenamefont {Hannam}\ \emph {et~al.}(2014)\citenamefont {Hannam},
  \citenamefont {Schmidt}, \citenamefont {Boh\'e}, \citenamefont {Haegel},
  \citenamefont {Husa}, \citenamefont {Ohme}, \citenamefont {Pratten},\ and\
  \citenamefont {P\"urrer}}]{PhysRevLett.113.151101}%
  \BibitemOpen
  \bibfield  {author} {\bibinfo {author} {\bibfnamefont {M.}~\bibnamefont
  {Hannam}}, \bibinfo {author} {\bibfnamefont {P.}~\bibnamefont {Schmidt}},
  \bibinfo {author} {\bibfnamefont {A.}~\bibnamefont {Boh\'e}}, \bibinfo
  {author} {\bibfnamefont {L.}~\bibnamefont {Haegel}}, \bibinfo {author}
  {\bibfnamefont {S.}~\bibnamefont {Husa}}, \bibinfo {author} {\bibfnamefont
  {F.}~\bibnamefont {Ohme}}, \bibinfo {author} {\bibfnamefont {G.}~\bibnamefont
  {Pratten}}, \ and\ \bibinfo {author} {\bibfnamefont {M.}~\bibnamefont
  {P\"urrer}},\ }\href {\doibase 10.1103/PhysRevLett.113.151101} {\bibfield
  {journal} {\bibinfo  {journal} {Phys. Rev. Lett.}\ }\textbf {\bibinfo
  {volume} {113}},\ \bibinfo {pages} {151101} (\bibinfo {year} {2014})},\
  \Eprint {http://arxiv.org/abs/1308.3271} {arXiv:1308.3271 [gr-qc]}
  \BibitemShut {NoStop}%
\bibitem [{\citenamefont {Foucart}(2012)}]{PhysRevD.86.124007}%
  \BibitemOpen
  \bibfield  {author} {\bibinfo {author} {\bibfnamefont {F.}~\bibnamefont
  {Foucart}},\ }\href {\doibase 10.1103/PhysRevD.86.124007} {\bibfield
  {journal} {\bibinfo  {journal} {Phys. Rev. D}\ }\textbf {\bibinfo {volume}
  {86}},\ \bibinfo {pages} {124007} (\bibinfo {year} {2012})},\ \Eprint
  {http://arxiv.org/abs/1207.6304} {arXiv:1207.6304 [astro-ph.HE]} \BibitemShut
  {NoStop}%
\bibitem [{\citenamefont {Pannarale}\ \emph {et~al.}(2011)\citenamefont
  {Pannarale}, \citenamefont {Rezzolla}, \citenamefont {Ohme},\ and\
  \citenamefont {Read}}]{Pannarale:2011pk}%
  \BibitemOpen
  \bibfield  {author} {\bibinfo {author} {\bibfnamefont {F.}~\bibnamefont
  {Pannarale}}, \bibinfo {author} {\bibfnamefont {L.}~\bibnamefont {Rezzolla}},
  \bibinfo {author} {\bibfnamefont {F.}~\bibnamefont {Ohme}}, \ and\ \bibinfo
  {author} {\bibfnamefont {J.~S.}\ \bibnamefont {Read}},\ }\href {\doibase
  10.1103/PhysRevD.84.104017} {\bibfield  {journal} {\bibinfo  {journal} {Phys.
  Rev. D}\ }\textbf {\bibinfo {volume} {84}},\ \bibinfo {pages} {104017}
  (\bibinfo {year} {2011})},\ \Eprint {http://arxiv.org/abs/1103.3526}
  {arXiv:1103.3526 [astro-ph.HE]} \BibitemShut {NoStop}%
\end{thebibliography}%
\bibliographystyle{apsrev4-1}

\end{document}